\documentclass[prd, preprintnumbers,amsmath,amssymb,nofootinbib
,superscriptaddress, twocolumn]{revtex4-2}

\usepackage{amsmath} 
\usepackage{amsthm} 
\usepackage{amssymb} 
\usepackage[hidelinks]{hyperref}
\usepackage{graphicx}
\usepackage{caption}

\usepackage{epsfig,subfigure,placeins,float} 
\usepackage{booktabs,ctable,multirow} 
\usepackage{exscale,relsize} 

\usepackage{subcaption}
\usepackage{siunitx}
\usepackage{lipsum}
\usepackage{parskip}
\usepackage{booktabs}
\usepackage{mathtools}
\usepackage[margin=0.6in]{geometry}
\usepackage [english]{babel}
\usepackage [autostyle, english = american]{csquotes}
\usepackage{float}

\usepackage{setspace}
\begin{document}
\title{Predicted Neutrino Signal Features of Core-Collapse Supernovae}    

\author{Lyla Choi}
\affiliation{Department of Physics, Princeton University, Princeton, NJ 08544}

\author{Adam Burrows}
\affiliation{Department of Astrophysical Sciences, Princeton University, Princeton, NJ 08544; Institute for Advanced Study, 1 Einstein Drive, Princeton, NJ 08540}

\author{David Vartanyan}
\affiliation{Carnegie Observatories, 813 Santa Barbara St., Pasadena, CA 91101}

\date{\today}
        
\begin{abstract}
In this paper, we examine the neutrino signals from 24 initially non-rotating, three-dimensional core-collapse supernova (CCSN) simulations carried to late times. We find that not only does the neutrino luminosity signal encode information about each stage of the CCSN process, but that the monotonic dependence of the luminosity peak height with compactness enables one to infer the progenitor core structure from the neutrino signal. We highlight a systematic relationship between the luminosity peak height with its timing. Additionally, we emphasize that the total energy radiated in neutrinos is monotonic with progenitor compactness, and that the mean neutrino energy contains a unique spiral SASI signature for nonexploding, BH-forming models. We also find that neutrino emissions are not isotropic and that the anisotropy increases roughly with progenitor compactness. To assess the detectability of these neutrino signal features, we provide examples of the event rates for our models for the JUNO, DUNE, SK, and IceCube detectors using the SNEWPY software \citep{snewpy}, and find that many of the trends in the luminosity signal can be detectable across several detectors and oscillation models. Finally, we discuss correlations between the radiated neutrino energy and the evolution of the gravitational-wave f-mode.    
\end{abstract}   

\maketitle

\section{Introduction}

At the end of its multi-million-year lifespan, the Chandrasekhar core of a massive star ($M > 8 M_{\odot}$) collapses, often triggering a violent supernova explosion. These supernova events are some of the most luminous and energetic events in the Universe, radiating more than $10^{53}$ erg of energy in neutrinos of all species \cite{Nagakura2020,Nagakura2021,Vartanyan_Burrows2023}, which offer unique insights into each stage of the core-collapse supernova (CCSN) process occurring deep within the star. 

The first and only such neutrino burst witnessed to date was detected by the Kamiokande-II and the Irvine-Michigan-Brookhaven (IMB) detectors in Japan and the United Sates, respectively, from the SN1987A supernova in the Large Magellanic Cloud. These detectors observed a total of 19 neutrino events over a period of around 13 seconds \cite{Bionta1987, Hirata1987, Scholberg2012, Janka2017}. While this detection enabled the verification of various aspects of the CCSN mechanism, including the formation and the Kelvin-Helmholtz cooling of the proto-neutron star (PNS), many aspects of supernova explosion theory have yet to be observationally verified, particularly the CCSN explosion mechanism itself \cite{Nagakura2020}. In the decades following SN1987A, there has been an increase in the number and sensitivity of neutrino detectors worldwide. These detectors include Super-Kamiokande (SK) \cite{Abe_2016} (a water Cherenkov detector that succeeded Kamiokande-II), the liquid-Argon detector DUNE \cite{Ankowski2016}, the upcoming liquid scintillator detector JUNO (Jiangmen Underground Neutrino Observatory) \cite{Abusleme_2025,An_2016}, and IceCube, a Cherenkov detector in Antarctica \cite{Abbasi2011, Kopke_2011}, among many others such as Hyper-Kamiokande \cite{HyperK2018}, HALO-2 \cite{Duba2008}, and RES-NOVA \cite{Alloni2025}, all of which open up a promising future for CCSN neutrino detection.

Overall, CCSN dynamics are governed by the competing effects of particle, nuclear, and gravitational physics. Given this computational complexity, as well as the relative infrequency of nearby and observable CCSN events, insight into the CCSN phenomenon has progressed largely through numerical simulations, aimed at determining the CCSN explosion mechanism and its correlated signatures \citep{Burrows2024}. Developments in computational power and efficiency have recently enabled our group using the code F{\sc{ornax}} \citep{Skinner2019} to calculate suites of longer-running three-dimensional CCSN simulations  \cite{Burrows2024,Burrows2023,Wang2024,Wang2024_nucleo}, that complement and extend shorter duration three-dimensional simulations carried out over the last decade \citep{muller2012,Tamborra2014, Couch_2015, Lentz2015,Muller2017,Ott2018,Summa2018,Glas2019, Muller2019_3D,Vartanyan_2019,Nagakura2020}. These new, longer-running 3D simulations are essential in order to capture 
complex phenomena missed in two-dimensional simulations, as well as key components of the neutrino lightcurve that manifest at late times.

Hence, one of the goals of our recent long-term 3D models is to provide a large set of neutrino signal templates to support the emerging international effort to characterize the next galactic supernova neutrino burst in light of modern detailed predictions. Our suite of 24 3D models includes systematic theoretical correlations between progenitor structure, supernova explosion characteristics, and remnant neutron star properties. They also incorporate the predicted correlations between progenitor compactness \citep{OConnor_2011}, explosion energy, mantle binding energy, nucleosynthetic yield \cite{Wang2024_nucleo}, gravitational-wave signals \cite{Vartanyan_2023, Choi2024}, and neutron star gravitational mass \cite{Burrows2024}. Most recently, these long-term, three-dimensional models have been used to study the highly asymmetric explosion and shock breakout of a progenitor evolved for up to five days post-bounce \cite{Vartanyan2024}, and to identify at least four distinct channels of black hole formation, each involving different explosion characteristics and neutrino signals, leading to black holes with different ranges of final mass \cite{Burrows2024_BH}.  

In this paper, we continue the work of providing predictions of key aspects of modern CCSN theory by highlighting the time- and species-dependent neutrino signatures of these 24 long-running three-dimensional simulations. In particular, we identify features in the predicted neutrino signals that vary systematically with progenitor star and compactness that are concrete theoretical predictions to be measured by the emerging international flotilla of underground and under-ice neutrino detectors. These include, but are not limited to, the systematics with progenitor core compactness and zero-age main-sequence mass (ZAMS) of the breakout burst, the time to the luminosity peak and its magnitude, the slope of the post-peak luminosity decay, and the evolution of the neutrino energy spectrum. We also discuss progenitor-dependent systematics of the total emitted neutrino energy, the degree and character of the emission anisotropy, and some correlations between the neutrino signal and the associated gravitational wave strain.  Though this is not our primary goal here, we also provide a few examples of the detectability of various features of the neutrino signal for the Super-Kamiokande (SK), IceCube, JUNO, and DUNE detectors using the SNEWPY software \cite{snewpy}. 

Much of the existing literature on detecting CCSN neutrinos has focused on distinguishing various groups' models despite their different simulation codes and progenitor masses \cite{Abe_2021,Mori_2022, Schnellbach2024}. While understanding the differences arising from the use of different simulation codes is a necessary exercise \cite{OConnor2018}, our goal with this study is to shift the focus of CCSN detection studies from distinguishing models with no physical basis for comparison to identifying general physical features and characteristics of CCSNe encoded in the predicted neutrino signals. These features should be the main scientific targets of those interested in supernova neutrino detection, falsifying supernova theory, and the underlying physics of core-collapse. 

This paper is organized as follows. In Section II, we provide an overview of our 24 models and the F{\sc{ornax}} simulation code. In Section III, we present each models' neutrino signals and discuss features of the luminosity and energy evolution of each species, as well as the angular anisotropy of the neutrino signal. We then convolve our simulation results with neutrino detector sensitivities using the SNEWPY software \cite{snewpy} in Section IV and discuss the detectability of various features of the neutrino signals. In Section V, we highlight correlations between the neutrino and the corresponding gravitational-wave signal, and summarize our findings and conclusions in Section VI. 

\section{Simulation Summary}
The 24 progenitor models studied in this paper were evolved in three dimensions with the radiation/hydrodynamic code F{\sc{ornax}} \cite{Skinner2019}. These models are initially non-rotating and range from 8.1 to 100 solar masses, thus spanning almost the entire spectrum of initial CCSN progenitor ZAMS masses. However, these models do not include other progenitor characteristics such as initial rotation or magnetic fields. These models have been studied in other contexts which analyzed the correlations between progenitor features and CCSN observables \cite{Burrows2024}, channels of BH formation \cite{Burrows2024_BH}, features of low-mass CCSN explosions \cite{Wang2024}, and the gravitational-wave (GW) signatures \cite{Choi2024}. The neutrino luminosity data for each model and species, as well as other aspects of the data can be found at \url{http://www.astro.princeton.edu/~burrows/nu-emissions.3d.update/} and \url{https://dvartany.github.io    /data/}. 

These simulations are the longest-running, 3D CCSN simulations to date, lasting over six seconds for some models, while previous 3D simulations generally ran for less than one second. The importance of longer simulations has become increasingly evident, as recent studies have revealed that strong temporal variations can emerge in neutrino signatures at late times \cite{Nagakura2021}.

\begin{table}[htpb!]
\begin{center}
\begin{tabular}{ |c|c|c|c|} 
 \hline
 \parbox{2.1cm}{\centering \vspace{0.2cm}Progenitor\vspace{0.2cm}}& Duration (s) & $\xi_{1.75}$ & \parbox{1.2cm}{NS/BH} \\ 
 \hline
 u8.1 & 0.84& 7.6 $\times 10^{-4}$ & NS \\ 
 9a & 1.78 & 6.7 $\times 10^{-5}$ & NS \\ 
 9b & 2.14 & 6.7 $\times 10^{-5}$ & NS\\
 9.25 & 3.53 & 2.5 $\times 10^{-3}$ & NS\\
 9.5 & 2.38 & 8.5 $\times 10^{-3}$ & NS\\
 z9.6 & 1.01 & 1.12 $\times 10^{-4}$ & NS \\
 11 & 4.49 & 0.12 & NS \\
 15.01 & 3.80 & 0.29 & NS\\
 16 & 4.18 & 0.35 & NS\\
 17 & 6.39 & 0.74 & NS\\
 18 & 8.51 & 0.37 & NS\\
 18.5 & 6.36 & 0.80 & NS\\
 19 & 7.00 & 0.48 & NS\\
 20 & 6.34 & 0.79 & NS\\
 21.68 & 1.57 & 0.84 & NS\\
 24 & 6.29 & 0.77 &NS\\
 25 & 6.32 & 0.80 & NS\\
 60 & 7.90 & 0.44 & NS\\
 \hline
 12.25 & 2.09 & 0.34 & BH* \\
 14 & 2.82 & 0.48 & BH*\\
 19.56 & 3.89 & 0.85 & BH \\
 23 & 6.23 & 0.74 & BH \\
 40 & 1.76 & 0.87 & BH\\
 100 & 0.44 & 1.02 & BH \\
 \hline
\end{tabular}
\end{center}
\caption{Summary of the 24 CCSN models evolved with the F{\sc{ornax}} code \cite{Skinner2019} indexed by their mass in solar mass units. ``u8.1'' and ``z9.6'' indicate progenitors of solar metallicity and zero metallicity, respectively. There are six BH-forming models, and the asterisk * denotes the BH-forming models which do not explode (12.25 and 14 $M_{\odot}$).}
\label{model_summary}
\end{table}

In Table \ref{model_summary}, we summarize the progenitor mass, duration of the simulation, compactness $\xi_{1.75}$\footnote{The compactness is defined as $\frac{M/M_\odot}{R(M)/1000\text{km}}$ \cite{OConnor_2011}. Here, we set $M$ equal to 1.75 $M_{\odot}$.}, and whether an NS or BH is formed. The varied durations reported here are a consequence of the models having been evolved on different machines and starting at different times during the recent years involved, which together result naturally in different end times. Some models were also stopped early if their dynamics had already naturally asymptoted. The label ``u8.1'' refers to a model of mass 8.1 $M_{\odot}$ with solar metallicity, and ``z9.6'' refers to a 9.6 $M_{\odot}$ progenitor with zero metallicity. Similarly, the ``9a'' and ``9b'' models both have a mass of 9 $M_{\odot}$, but the former has imposed perturbations from convection. Six of our models (12.25, 14, 19.56, 23, 40, and 100) collapse into BHs during or after the end of the simulation, with the 12.25 and 14 models collapsing into BHs without exploding. A more detailed investigation into the channels of BH formation for these models can be found in  \citet{Burrows2024_BH}.

\section{Neutrino Signal Results}
\subsection{Neutrino Luminosities}
\begin{figure*}[htbp]
   \centering
    \includegraphics[width=0.47\linewidth]{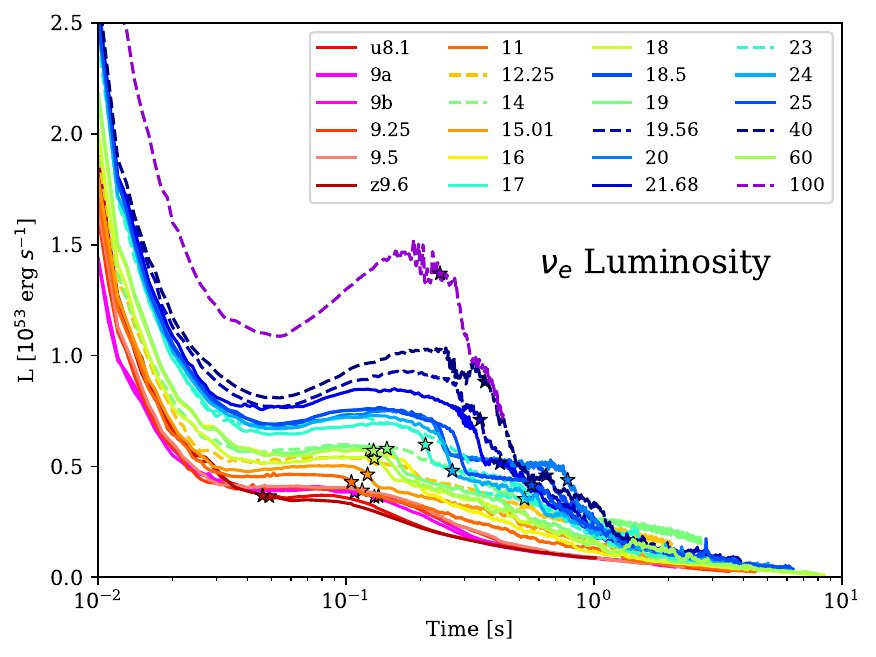}
    \includegraphics[width=0.47\linewidth]{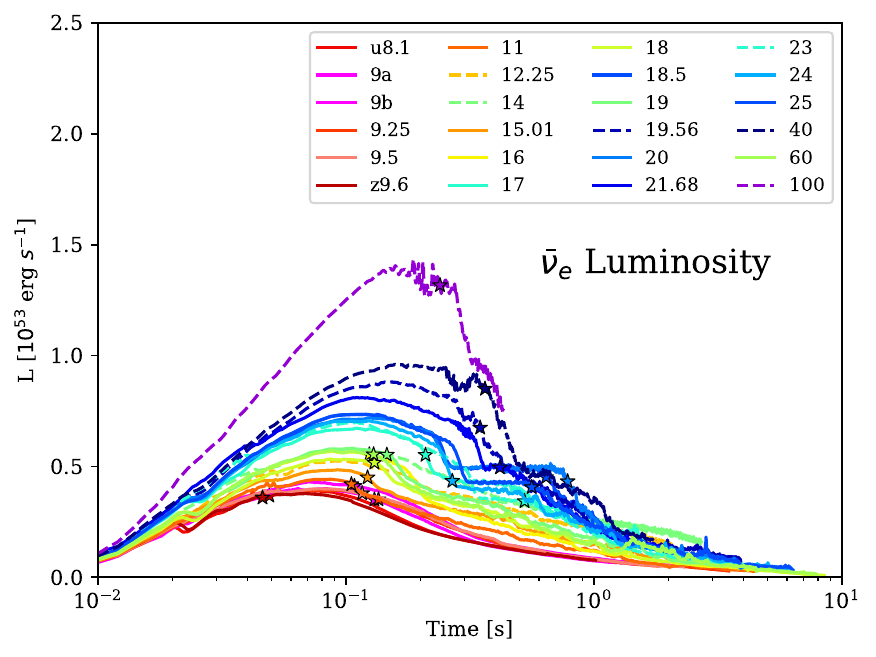}
    \includegraphics[width=0.47\linewidth]{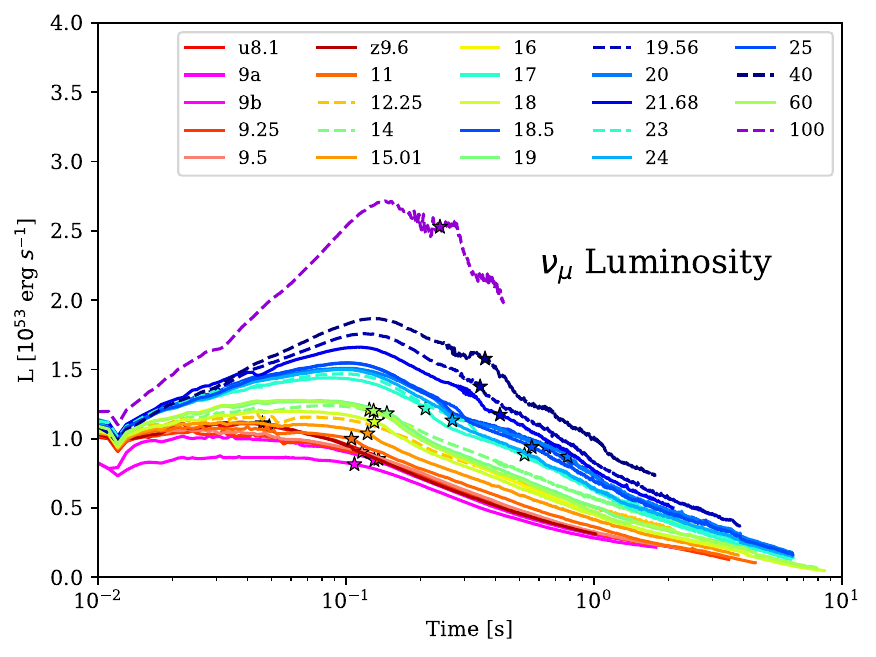}
    \captionsetup{justification=raggedright}
    \caption{Angle-averaged luminosities for each neutrino species of the 24 models colored by compactness and plotted in a log time scale (in seconds). The BH-forming models are plotted with dotted lines, and the explosion time for each model, defined as the time when the shock radius grows larger than 180 kilometers, is plotted with a star. The peak luminosity varies by around a factor of four between the most compact and least compact models.  In the $\nu_e$ signal in particular, we observe each major ``stage'' of the CCSN process, including 1) the decay from the breakout burst; 2) the trough at around $\sim$0.1 seconds; 3) the rise to a peak due to accretion; and 4) the decay from that plateau after explosion (for most models), associated in part with the concomitant decrease in accretion and the subsequent Kelvin-Helmholtz cooling phase of the residual proto-neutron star. After a black hole forms (if it does), the neutrino signals stop abruptly. The $\bar{\nu}_e$ and $\nu_{\mu}$ species display the rise to the peak, followed by the cooling and explosion-diminished accretion phase. While the u8.1 and z9.6 models explode before the luminosity signal peaks in all species, for the other models, the explosion occurs towards the end of the accretion phase of the signal. For non-exploding, ``Channel 4'' BH-forming models \citep{Burrows2024_BH}, there is a distinguishing high-frequency behavior at late times ($>$ 1 second) due to the spiral SASI. These plots indicate a monotonic dependence on compactness for all species' luminosities (the models are cleanly nested), with the more compact progenitors having greater luminosities and the less-compact models having shallower and smoother curves. This compactness-dependent behavior enables us to infer properties of the progenitor core and the progress of the supernova explosion directly and in ``real time" from neutrino observations.
    \label{luminosities24}}
    \end{figure*}

The luminosities for all 24 models and each neutrino species are plotted in logarithmic time in Figure \ref{luminosities24}. In this study, we distinguish between the electron neutrino $\nu_e$, electon anti-neutrino $\bar{\nu}_e$, and bundle all other ``heavy'' neutrinos and their anti-neutrinos as ``$\nu_{\mu}$''. These luminosities are averaged across all viewing angles and are colored by compactness. The explosion times are also plotted, defined as the time when the shock radius grows larger than 180 kilometers. 

From the overall shape of each species' luminosity curves, we can follow each stage of the CCSN process. In the $\nu_e$ luminosity curve in particular, we observe the decline from the initial breakout burst within the first $\sim$0.1 seconds, a gradual rise to a plateau, and the decay from the plateau during a (often post-explosion) Kelvin-Helmholtz cooling of the PNS at late times. For the $\bar{\nu}_e$ and $\nu_{\mu}$ luminosities, we do not observe a breakout burst in the first 0.1 seconds, as the burst largely occurs due to electron capture on protons which produces $\nu_e$ neutrinos. However, the other species still display a gradual rise to a plateau as material accretes onto the PNS and the electron lepton number excess around the neutrinospheres abates. Additionally, the majority of the models undergo an explosion by the end of the accretion stage of the luminosity signal, with the u8.1 and z9.6 models being the exceptions due to their prompt explosion. Finally, we observe BH formation via the ``Channel 1'' and ``Channel 3'' modalities \citep{Burrows2024_BH} in the behavior of all species' light curves due to the abrupt end to the 19.56, 40, and 100 $M_{\odot}$ signals. We also note that the three highest-luminosity models of all these models across all species all form BHs. 

At later times (after around one second), the non-exploding BH-forming models (12.25 and 14 $M_{\odot}$) exhibit sustained, high-frequency oscillations in the $\nu_e$ and $\bar{\nu}_e$ luminosity signals, remaining relatively constant at around $0.3 \times 10^{53}$ erg s$^{-1}$, while the luminosities of the other models decay. Taking a Fourier-transform of this component of the 12.25 and 14 models' luminosities and comparing them to the spectra of the corresponding GW data \cite{Choi2024}, both signals display a soft peak or ``bump'' near 100 Hz and below, which has been attributed to the spiral SASI behavior \cite{Nagakura2020, Shibagaki2021, Vartanyan_2023}, in which the stalled shock wave spirals in a rotating low-order mode. However, we leave a more detailed discussion of the frequency content of both signals for future work. The 12.25 and 14 models are classified in \citet{Burrows2024_BH} as representatives a unique category of ``Channel 4'' of BH-formers that do not explode. Therefore, the late-time sustained, high-frequency luminosity behavior may serve as an observational signature of this channel of BH formation. 

From the color scheme of Figure \ref{luminosities24}, we can see that for all neutrino species, the magnitude of the luminosity is monotonic with compactness, a result consistent with previous studies \cite{OConnor_2013, Vartanyan_Burrows2023}. For each species, the peak luminosity varies by around a factor of three to four between the least- and most-compact models, from around $0.37 \times 10^{53}$ erg s$^{-1}$ to $1.53 \times 10^{53}$ erg s$^{-1}$ for the $\nu_e$ luminosities and from around $0.9 \times 10^{53}$ erg s$^{-1}$ to $2.7 \times 10^{53}$ erg s$^{-1}$ for the $\nu_{\mu}$ luminosity. When considering just the NS-forming models, the peak luminosity values still vary by a factor of two between the most and least compact models for all neutrino species. In general, the compactness parameter serves as an indicator of the progenitor density profile, and is thus associated with the accretion rate $\dot{M}$, as denser, more-compact progenitors will accrete at higher rates \cite{OConnor_2013}. Therefore, this compactness-dependent nesting in the luminosity is expected, as more compact models have higher accretion-powered luminosities. Similarly, we observe that the width (or duration) of the accretion luminosity plateau also increases with compactness, indicating that more compact progenitors have higher accretion rates and accrete for longer than their less-compact counterparts. These compactness-dependent characterizations of the luminosity curves, especially observed over a set of 24 models, enables us to identify and distinguish systematic behaviors associated with the structure of the progenitor Chandrasekhar core that can be probed with CCSN neutrino detectors. For example, the observation of $\nu_e$ or $\bar{\nu}_e$ lightcurves attaining luminosities as high as $1.5 \times 10^{53}$ erg s$^{-1}$ would suggest a high progenitor compactness of $\xi_{1.75} \sim 1$, while observing a $\nu_e$ or $\bar{\nu}_e$ luminosity peak of $\sim 0.4 \times 10^{53}$ erg s$^{-1}$ likely indicates a low-compactness progenitor with $\xi_{1.75}$ on the order of $10^{-4}-10^{-3}$.

\begin{figure}[htbp]
    \centering
    \includegraphics[width=1\linewidth]{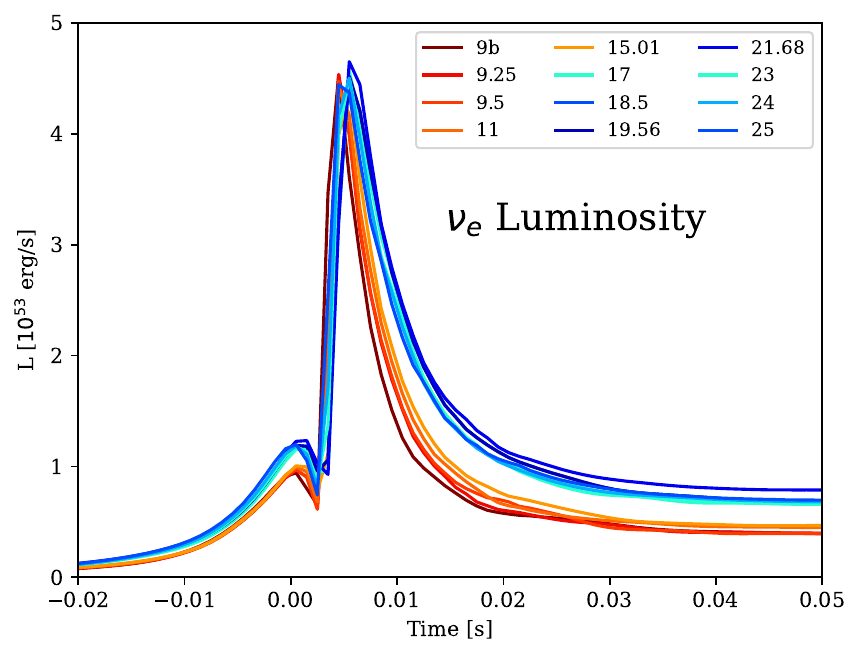}
    \captionsetup{justification=raggedright}
    \caption{The breakout burst of the $\nu_e$ neutrinos over a period of 0.07 seconds for 12 of the models colored by compactness. The height of the breakout burst is approximately independent of progenitor mass and compactness, but the width of the breakout burst increases with compactness, with a full-width at half maximum ranging from $\sim$6 to $\sim$10 milliseconds. This trend is due to the monotonically increasing dependence of mass accretion rate after bounce on compactness. There is also a smaller precursor peak at around zero seconds due to the neutronization of the collapsing core before the main breakout burst that has a slight dependence on compactness.}
    \label{breakout_burst}
\end{figure}

In Figure \ref{breakout_burst}, we plot the $\nu_e$ breakout burst in linear time for 12 models. This phase of the simulations was evolved in two dimensions and was part of a larger suite discussed in a previous work \cite{Vartanyan_2023}. The breakout burst is a feature solely manifest with the $\nu_e$ neutrinos and is caused by electron capture on shock-liberated free protons associated with the shock's generation and initial outward progress. We observe that the peak of the breakout burst rises to around $4.5 \times 10^{53}$ erg s$^{-1}$, dwarfing the later features of the $\nu_e$ signal which peak around $\sim$0.5-1.5 $\times 10^{53}$ erg s$^{-1}$. Unlike the other aspects of the $\nu_e$ luminosity signal, the height of this breakout peak is roughly the same for all 12 models, regardless of progenitor mass or compactness. The compactness becomes relevant only shortly after the signal peaks, with less-compact models having faster and steeper decays compared to the more-compact models, leading to less-compact models with narrower breakout peaks and more compact models with wider peaks. Each model also experiences an earlier, smaller peak at around ``zero" seconds due to the neutronization of the collapsing core and subsequent neutrino trapping \cite{Thompson_2003, Wallace_2016}. The height of this pre-breakout mini-peak appears to have a slight dependence on compactness, with a slightly higher peak for the more compact models. This slight dependence on progenitor structure may serve as a physical target for studies interested in resolving the CCSN bounce within 10 milliseconds \citep{Halzen2009,Pagliaroli2009}. 

\begin{figure*}[htbp]
    \centering
    \includegraphics[width=0.47\linewidth]{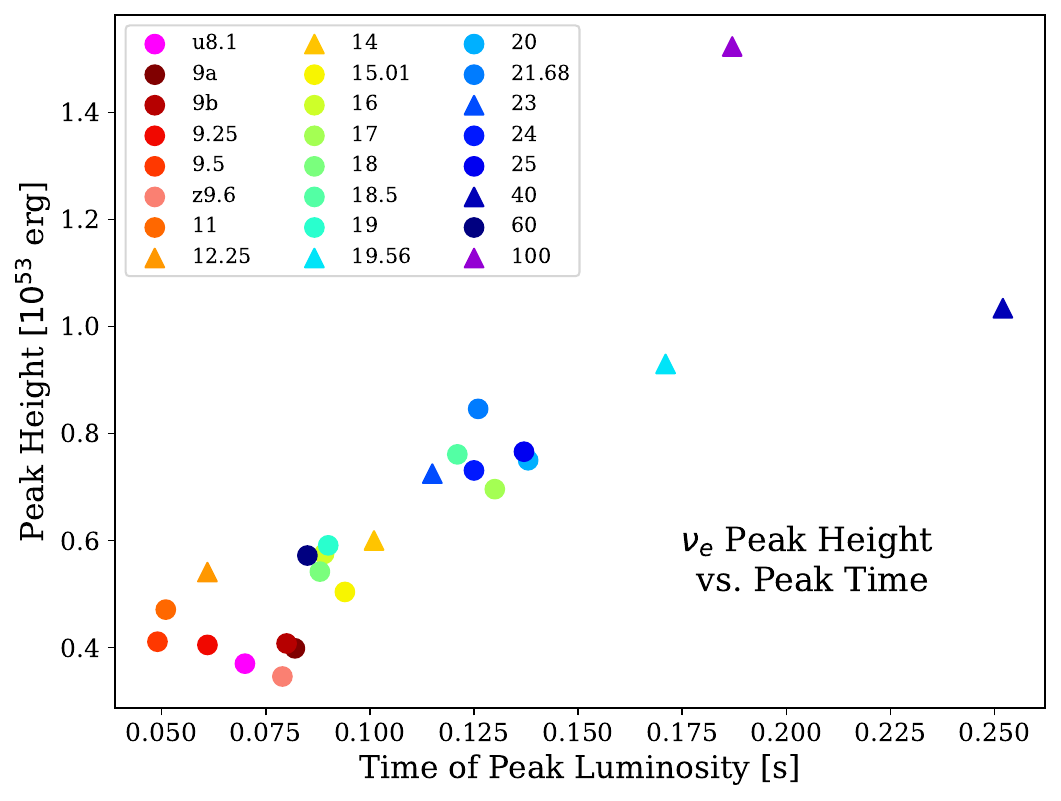}
    \includegraphics[width=0.47\linewidth]{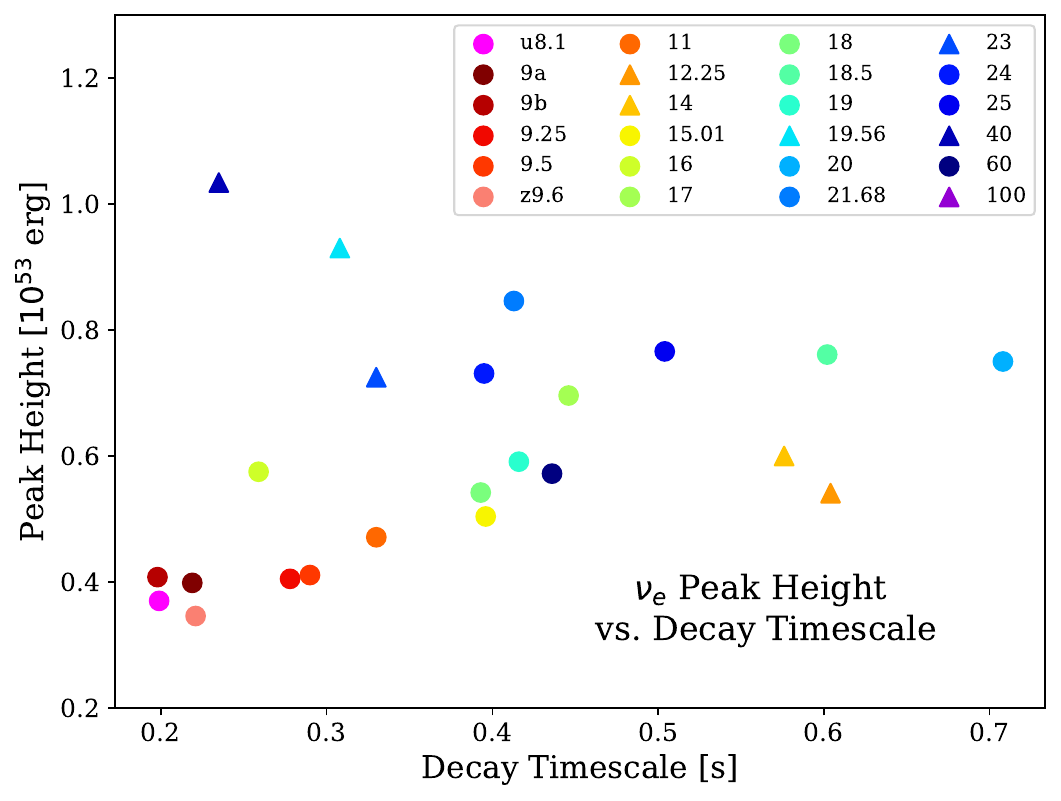}
    \includegraphics[width=0.47\linewidth]{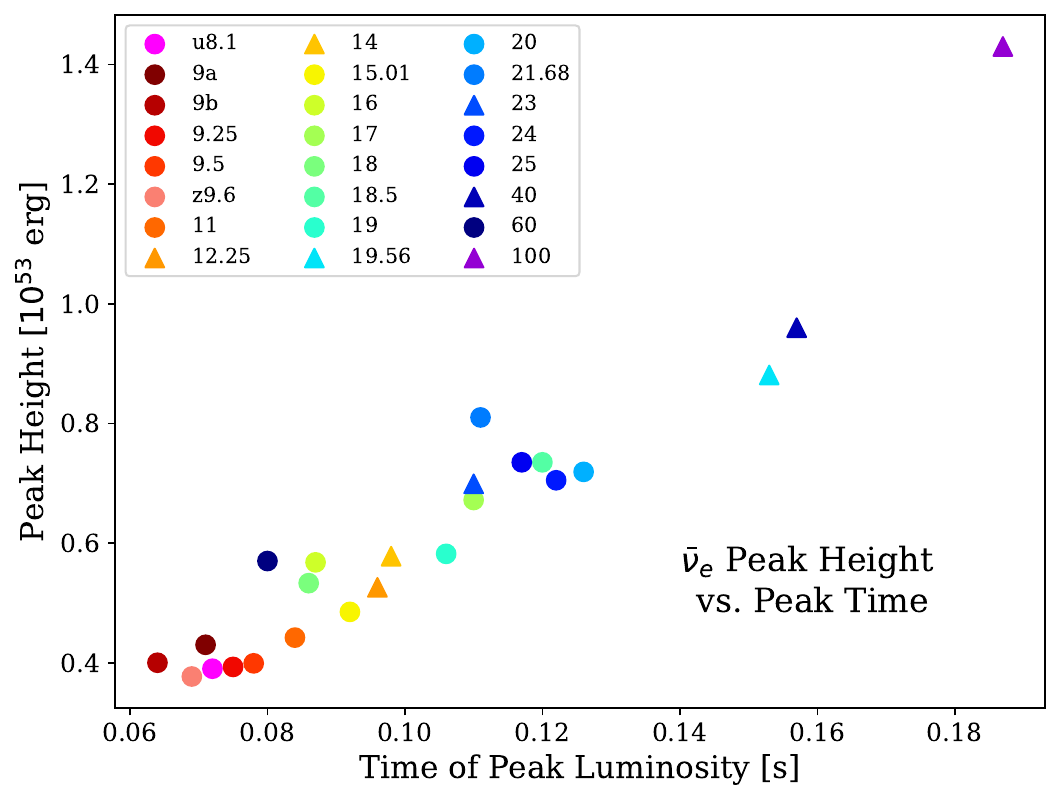}
    \includegraphics[width=0.47\linewidth]{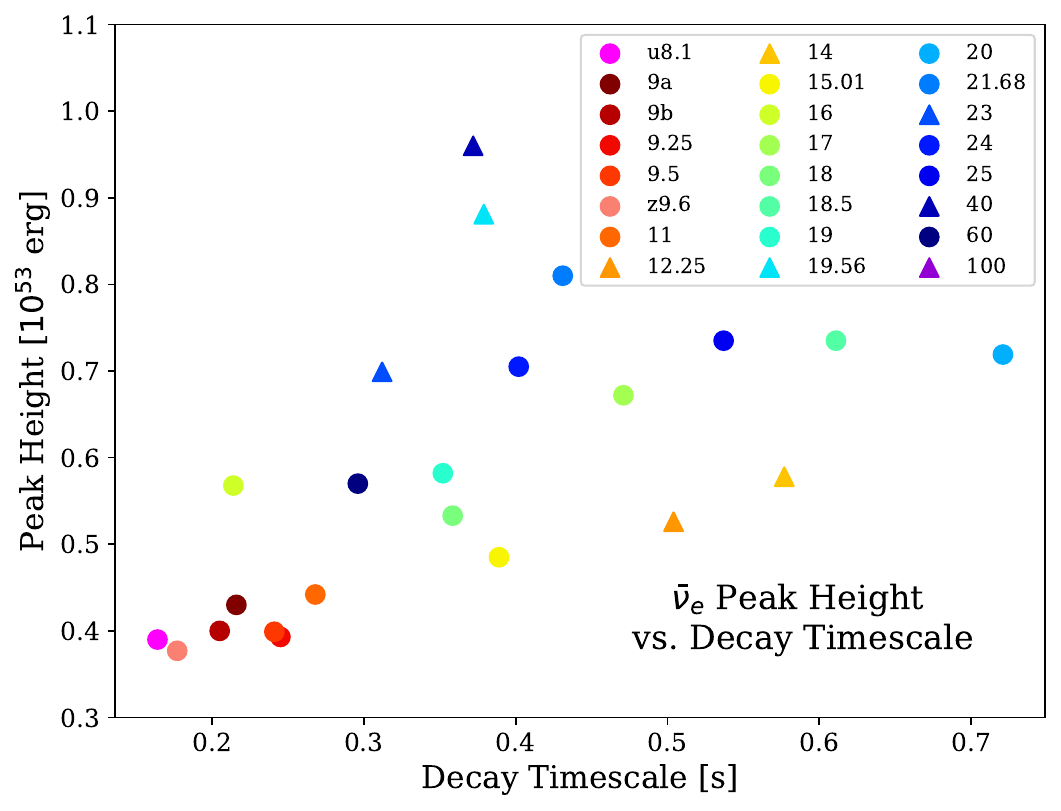}
    \includegraphics[width=0.47\linewidth]{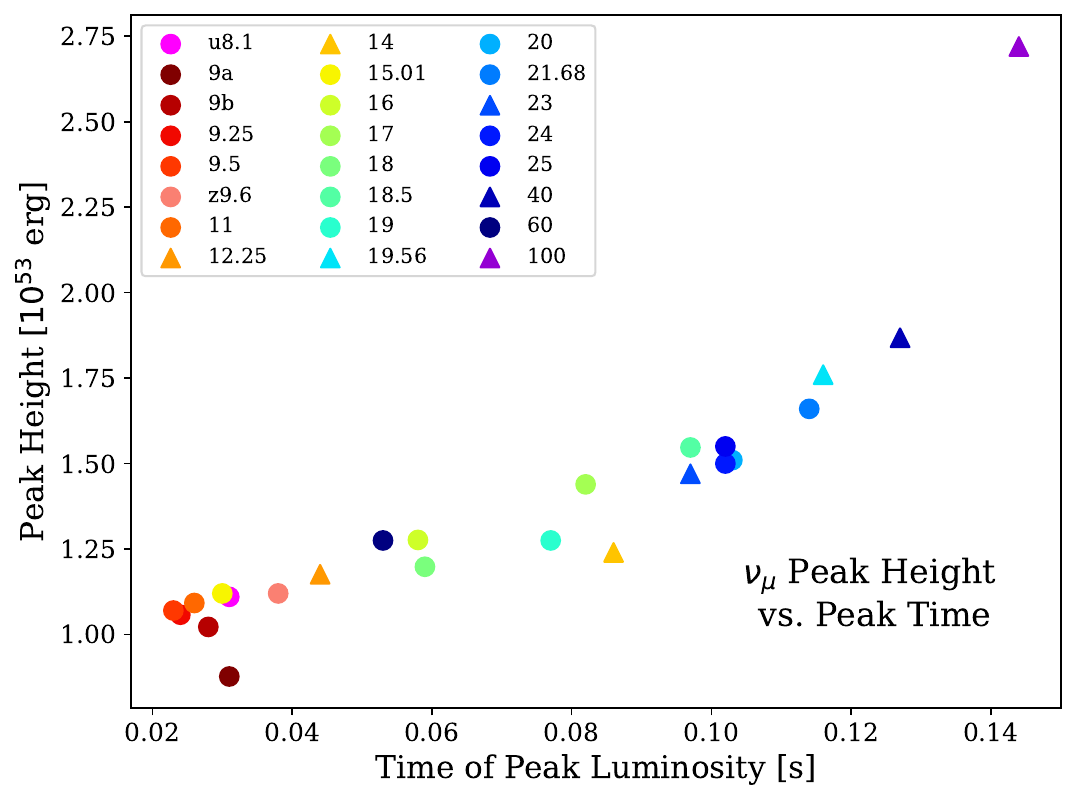}
    \includegraphics[width=0.47\linewidth]{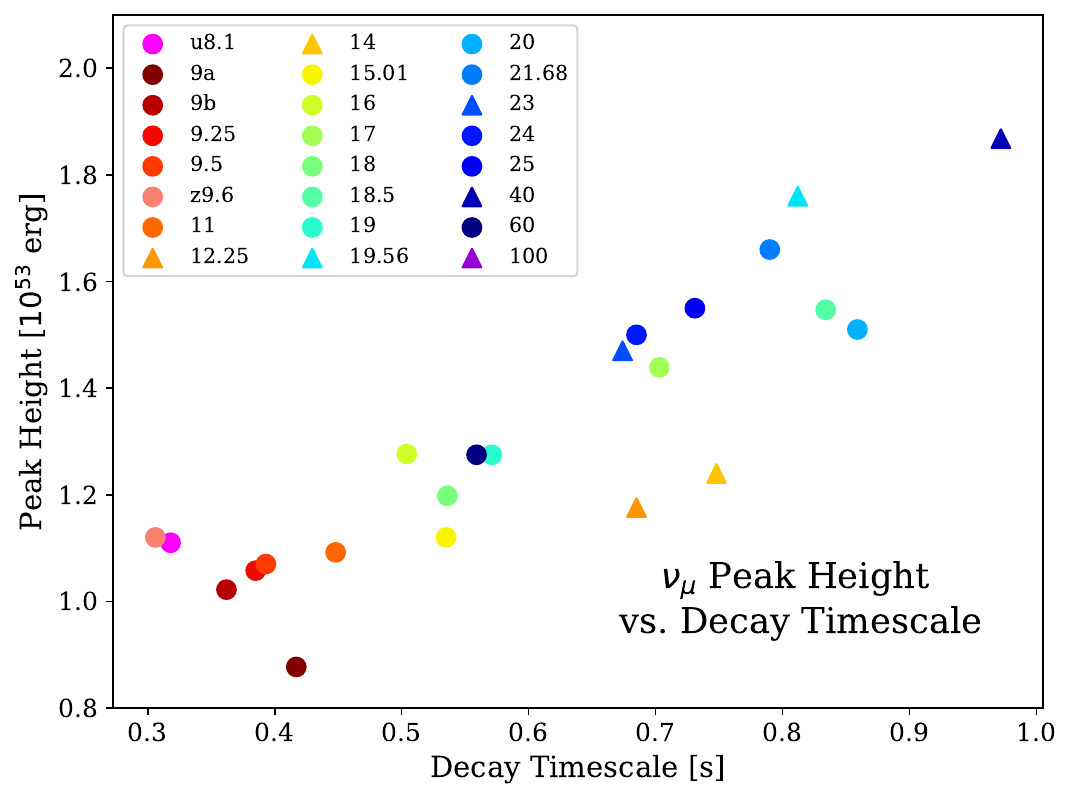}
    \captionsetup{justification=raggedright}
    \caption{Left column: Peak luminosity (after the breakout burst) attained for all neutrino species as a function of the time of the peak, for all models. Right column: Peak luminosity attained for all species as a function of the decay timescale from the peak in seconds, calculated from the time elapsed before the luminosity reaches half of the peak value. The BH-formers are plotted with triangles. Across all species and particularly for the NS-formers, there is a characteristic distribution of neutrino luminosity peak, timing, and width that may serve as a key diagnostic of CCSN theory. This includes a clear, monotonic relationship between the maximum luminosity attained and the time at which this peak occurs, with higher peaks from higher-mass progenitors occurring later. The trend between peak luminosity and the duration of the peak in seconds is slightly weaker, particularly for the $\bar{\nu}_e$ neutrinos, but is much stronger for the $\nu_{\mu}$ neutrinos.}
    \label{lum_height_time}
\end{figure*}

To better understand the systematics of the luminosity peaks and decay timescales for all models, in Figure \ref{lum_height_time} we plot for each species the peak luminosity attained (during the plateau phase, after the initial breakout burst) as a function of the time of the peak (left column) and the decay timescale of the peak (right column). Overall, we observe a strong, monotonic relationship between the peak height and the time at which this peak occurs across all species, with the higher luminosities peaking at later times. The latest-time, highest-luminosity models, specifically the 19.56, 40, and 100 $M_{\odot}$ models, all form BHs. The earlier, lower-luminosity peaks are from the lowest-mass and compactness progenitors in our suite. In the peak height versus decay timescale plots, there is an overall monotonic relationship that is most explicit for the $\nu_{\mu}$ neutrinos and more ambiguous for the $\bar{\nu}_e$ neutrinos. With this approximate relationship, we observe that lower-mass progenitors have lower-luminosity peaks that last for a shorter time compared to higher-mass progenitors. The outliers in this trend for the $\nu_e$ and $\bar{\nu}_e$ neutrinos are generally the BH-formers. This is because lower-mass models generally have lower accretion rates that result in lower luminosities and shorter timescales, whereas higher-mass models have higher accretion rates that lead to larger luminosity peaks and longer decay timescales. However, lower-mass, BH-forming models that do not explode and continue to accrete necessarily have lower peak luminosities (due to the low accretion rate), but still long decay timescales (as the accretion is prolonged), causing them to fall out of this trend between peak luminosity and decay timescale. Whether non-exploding BH-forming models should actually be in this low mass range has yet to be determined \citep{Burrows2024_BH}. 

In addition to characterizing the luminosity peak height and timing for each species, we calculated the slope of the rise to the peak luminosity (using the linear form $L \sim m\ln t + b$) and the decay from the peak (using the power law form $L \sim At^a$), employing a curve-fitting algorithm to find the parameters $m$, $b,$ $A,$ and $a$ for each model and neutrino species. Across all models and species, the magnitude of the slope of the linear rise to the peak as well as the power-law decay from the peak are roughly monotonic with progenitor compactness. In particular, the decay from the peak $\nu_e$ luminosity ranges from $L_{\nu_e} \sim t^{-0.62}$ to $t^{-1.17}$ between the least- and most-compact progenitors, excluding the non-exploding 12.25 and 14 models which exhibit late-time spiral SASI behavior distinct from a power-law decay. Similarly, the decay from the peak $\bar{\nu}_e$ luminosity ranges from $L_{\bar{\nu}_e} \sim t^{-0.62}$ to $t^{-1.06}$ between all of the exploding models. When fitting the slope of the rise to the peak luminosity, we found that the $\bar{\nu}_e$ slopes range from $L_{\bar{\nu}_e} \sim0.16\ln t$ to $\sim0.53 \ln t$ between the least- and most-compact progenitors. For the $\nu_{\mu}$ neutrinos, the rise in luminosity goes as $L_{\bar{\nu}_{\mu}} \sim0.054\ln t$ to $\sim0.63 \ln t$, depending on the model. 

Overall, these plots reveal distinct correlations between peak neutrino luminosities, time to peak, width of peak, and luminosity rise and decay behavior for almost all of the models in our suite of simulations. Such a characteristic distribution of luminosity peak height, duration, width, slope, and decay is a prediction of our supernova theory simulations that has rarely before been highlighted and serves as a key prediction of modern CCSN theory.

\begin{figure*}[htbp!]
    \centering
    \includegraphics[width=0.49\linewidth]{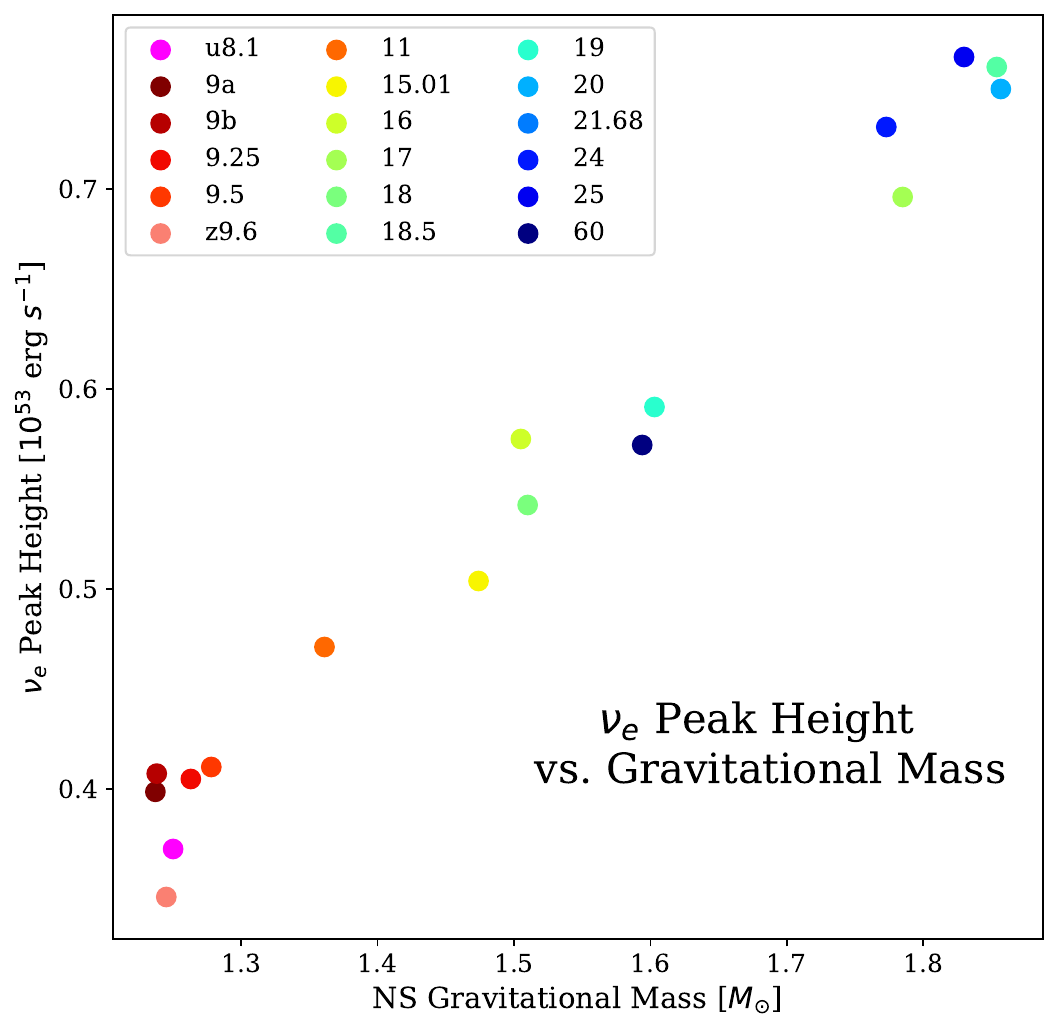}
    \includegraphics[width=0.49\linewidth]{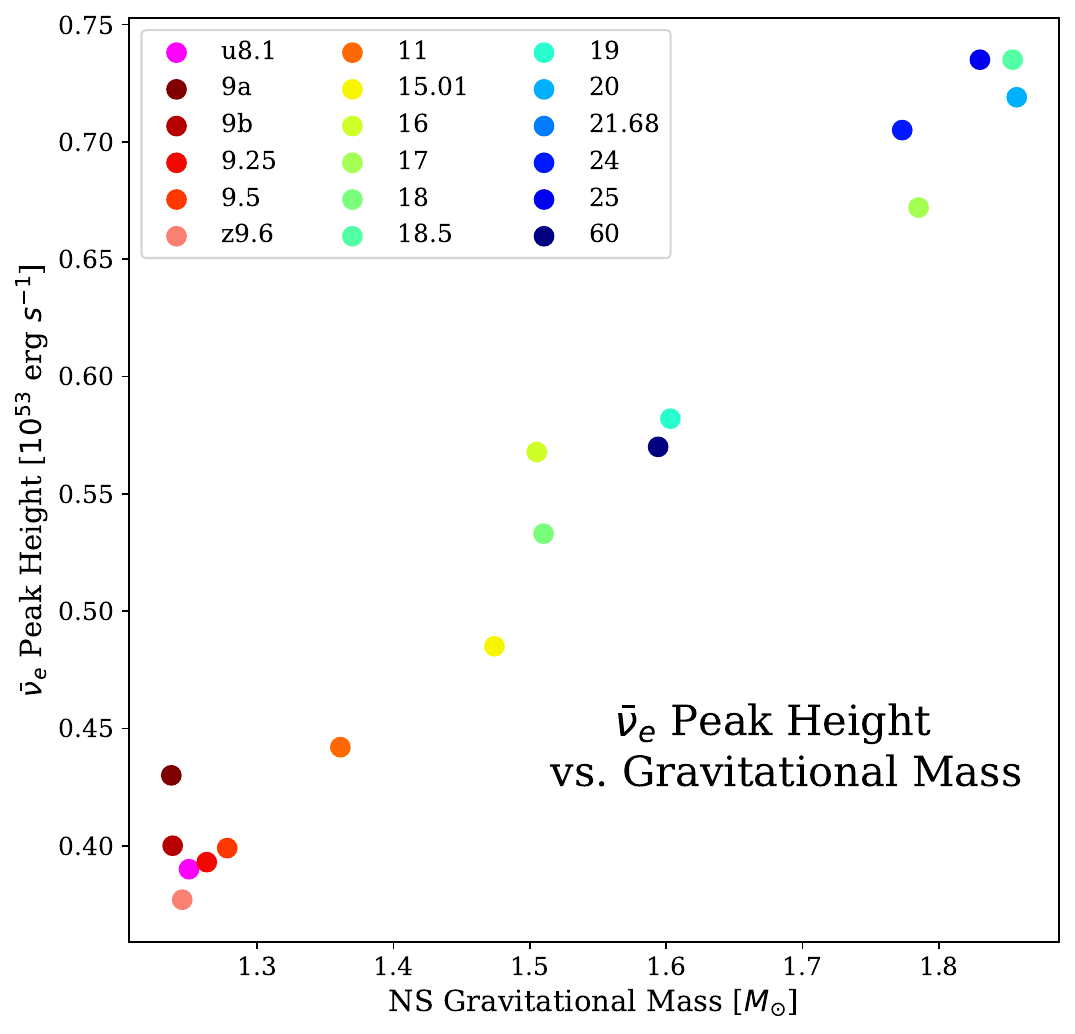}
    \includegraphics[width=0.49\linewidth]{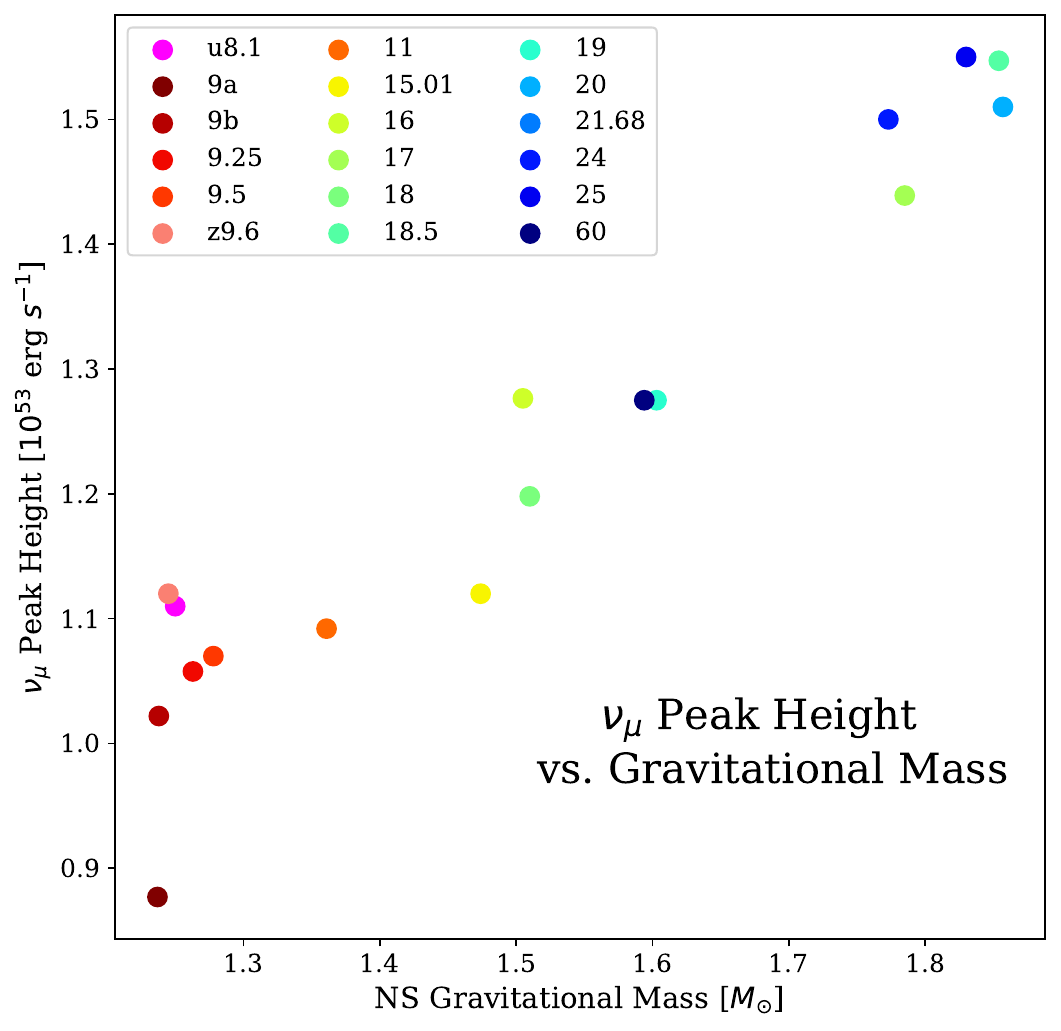}
    \includegraphics[width=0.49\linewidth]{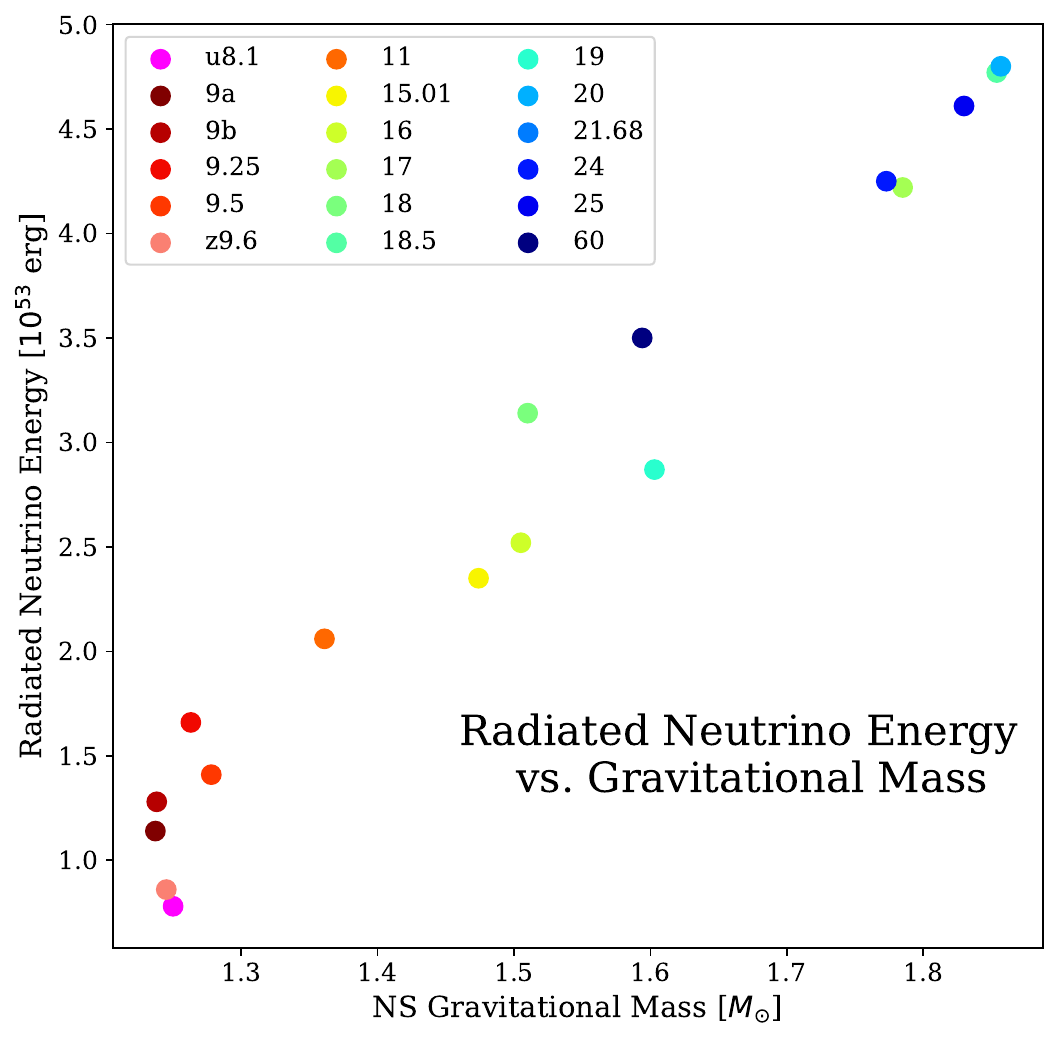}
    \captionsetup{justification=raggedright}
\caption{Correlations between peak luminosity attained for each species and the total energy radiated via neutrino emission as a function of final NS gravitational mass. The monotonic correlation in all plots for models $> 11 M_{\odot}$ enables us to estimate the NS gravitational mass from peak and total neutrino luminosity measurements. \label{correlations}}
    \end{figure*}

In Figure \ref{correlations}, we plot the peak luminosity attained for each neutrino species, as well as the total neutrino energy radiated during each simulation as a function of the gravitational mass of the residual NS for the NS-forming models. While there is some degeneracy for progenitors around 9 $M_{\odot}$, each plot displays a clear, monotonic behavior with gravitational mass for progenitor models $\sim 11\ M_{\odot}$ or greater, thus enabling us to infer properties of the NS just from the neutrino signal of intermediate to high-mass progenitors, given the derived theoretical context. 

In summary, the observation of the luminosity curves of each species by neutrino detectors enables us to witness in real time the stages through which the core progresses during the supernova phenomenon, as well as the systematics with progenitor compactness and residual neutron star mass. The main features in the neutrino signal to identify and characterize include:
\begin{enumerate}
    \item The overall shape of the neutrino light curves including 1) a breakout burst just before 0.01 seconds after bounce; 2) a more gradual rise in the $\nu_e$ signal starting at 0.1 seconds; 3) a plateau; and 4) a decay in the signal at later times, stages which have been highlighted in previous CCSN neutrino luminosity studies as well \cite{Hudepohl2010, Kuroda2012,Muller_2014,Wallace_2016}. 
    \item The overall height and duration of the luminosity peak for all species, which enables us to constrain progenitor compactness due to the monotonic nesting of peak height as a function of compactness. The peak height also enables us to infer properties of the residual NS, such as its gravitational mass.
    \item The observation of sustained, late-time, high-frequency oscillations that may enable us to identify Channel 4 \citep{Burrows2024_BH} BH formation. 
    \item The characteristic relationship between the height of the luminosity peak for each species and the time at which this peak occurs. 
\end{enumerate}

\subsection{Total Emitted Neutrino Energy}

\begin{figure*}[htbp]
    \centering
    \includegraphics[width=0.48\linewidth]{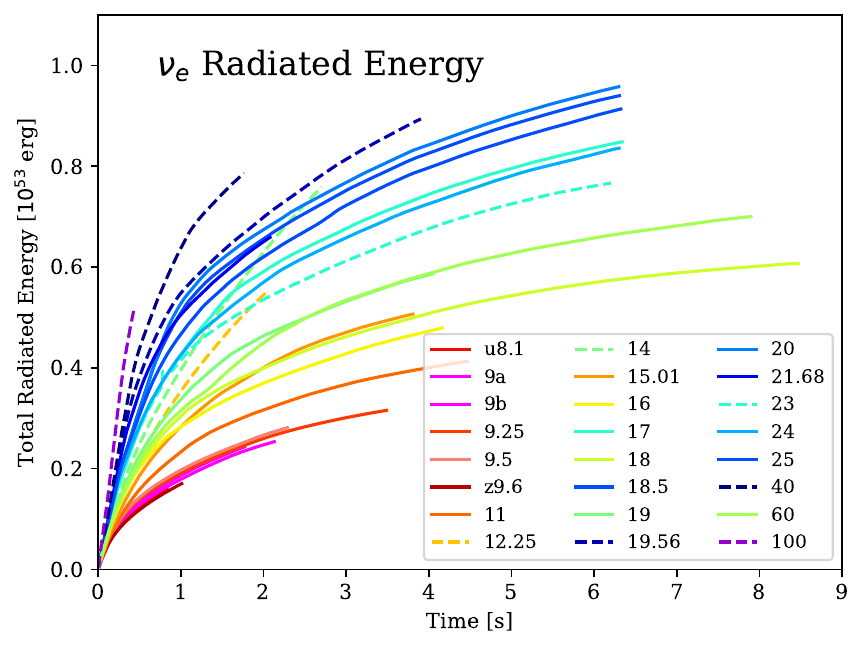}
    \includegraphics[width=0.48\linewidth]{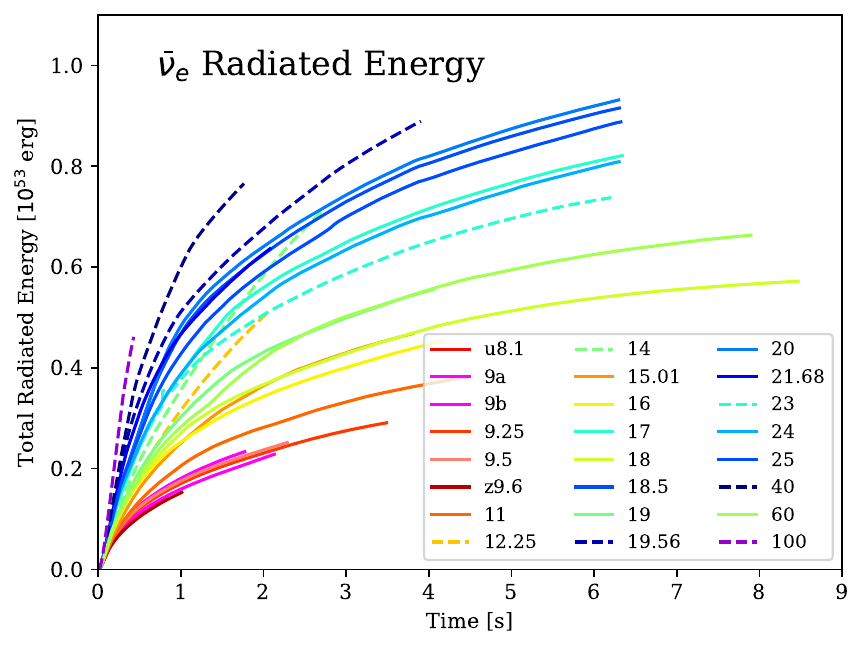}
    \includegraphics[width=0.48\linewidth]{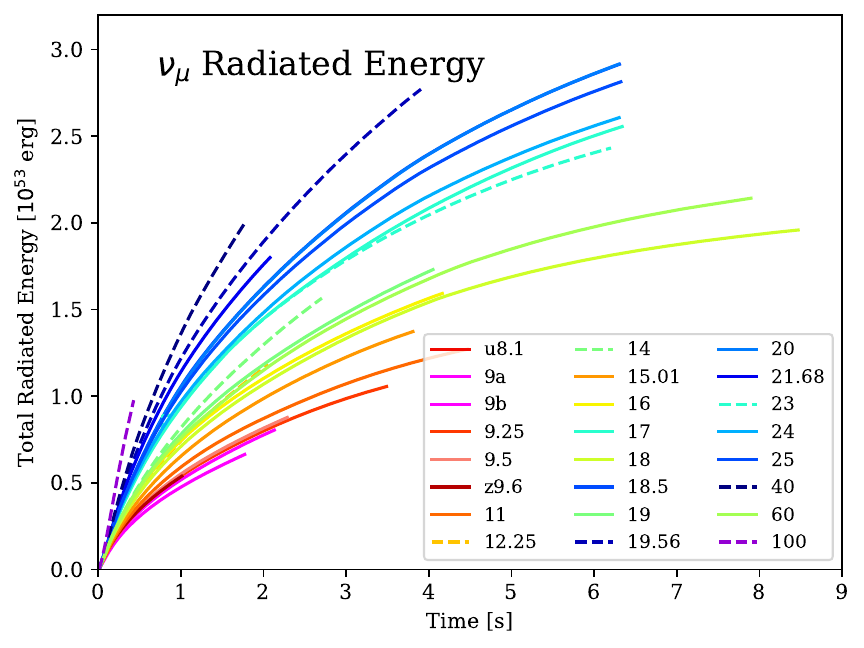}
    \includegraphics[width=0.48\linewidth]{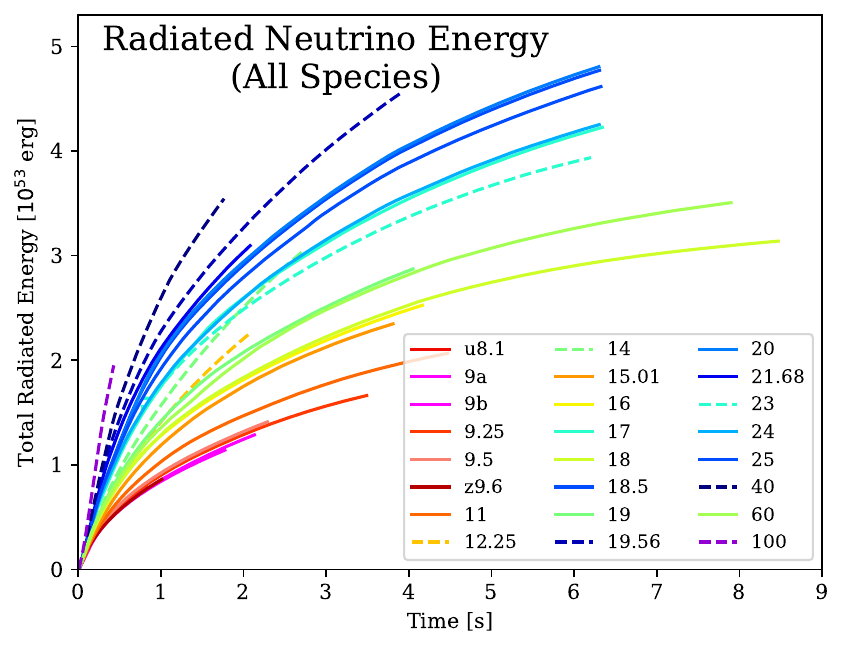}
    \captionsetup{justification=raggedright}
    \caption{Radiated neutrino energy averaged across all viewing angles for each model and neutrino species colored by compactness as a function of time. The BH-forming models are plotted with dotted lines. The energies of all models increase until the end of the simulation, implying the continued loss of neutrinos and motivating the need for longer simulations. As with the luminosity signals, we observe a monotonic dependence of neutrino energy of all species with compactness, enabling us to infer progenitor core structure from neutrino losses, as a signal radiated $2 \times 10^{53}$ erg in 0.5 seconds likely indicates a very compact ($\xi\sim 1$) model, whereas observed neutrino losses of $0.5 \times 10^{53}$ erg implies a compactness on the order of $10^{-4} - 10^{-3}$. Finally, we note that the three models with the largest neutrino energies all form BHs. \label{tot_energy}}
    \end{figure*}

In addition to the luminosity curves, the total radiated neutrino energy contains a wealth of information concerning CCSN stages and progenitors. In Figure \ref{tot_energy}, we plot the total radiated neutrino energy for each species and model as a function of time and colored by compactness. This ``total energy'' is defined by integrating the angle-averaged luminosities of each species from Figure \ref{luminosities24} and, for the bottom row, summing over each species. Overall, we can see that while the slopes for all models and species are steepest in the first $\sim$0.5 seconds, the total energy is still increasing by the end of the simulation for all models. Such behavior indicates that the models continue to lose energy via neutrino emission even after the simulation ends, further motivating the need for even longer simulations. From the color scheme of Figure \ref{tot_energy}, we can see that as with luminosity signals, the total energy radiated via neutrinos is monotonic with compactness, with more compact models corresponding to larger neutrino energy losses. Once again, this compactness-dependent nesting enables the use of total neutrino energy evolution to constrain progenitor core structure. A signal that radiates $\sim2 \times 10^{53}$ erg within the first 0.5 seconds is likely to have a very compact progenitor with $\xi_{1.75} \sim 1$, but if the radiated energy is only $\sim0.5 \times 10^{53}$ erg within the same time span, its compactness may be on the order of $10^{-4}-10^{-3}$ instead. We also note that the three most-compact models that radiate the most energy in neutrinos all eventually collapse into BHs, and thus, both total neutrino energy and abrupt signal cutoff serve as indicators of BH formation. 

When comparing the percent contribution of each neutrino species to the total neutrino energy radiated by the end of the simulation, we observe that for the NS-forming models, the total radiated energy is dominated by the sum of the heavy neutrino species ``$\nu_{\mu}$'', with the $\nu_e$, $\bar{\nu}_e$, and ``$\nu_{\mu}$'' species contributing around 20\%, 19\% and 61\%, respectively. The BH-forming models, however, (in particular the 12.25, 14, 40, and 100 $M_{\odot}$) have noticeably higher relative $\nu_e$ and $\bar{\nu}_e$ contributions compared to NS models at around 25\% and 23\%, respectively, bringing the percent contribution of the heavy $\nu_{\mu}$ species  down to around 52\%. However, this trend is not seen for the 19.56 and 23 $M_{\odot}$ BH models and is more subtle for the 40 $M_{\odot}$ model. Using the BH formation channels identified in an earlier work \cite{Burrows2024_BH}, these results indicate that Channels 3 and 4 have the higher electron-type and lower heavy ``$\nu_{\mu}$''-type neutrino contributions compared to NS-forming models. Therefore, the relative neutrino species contribution to the total energy radiated in neutrinos may serve as a weak indicator of BH formation.

\begin{figure*}[htbp]
    \centering
    \includegraphics[width=0.49\linewidth]{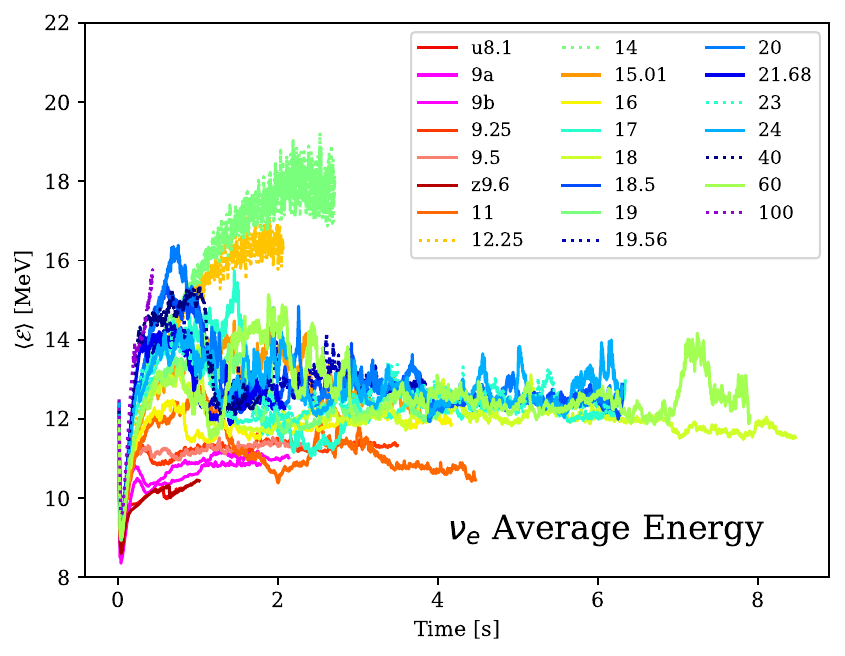}
    \includegraphics[width=0.49\linewidth]{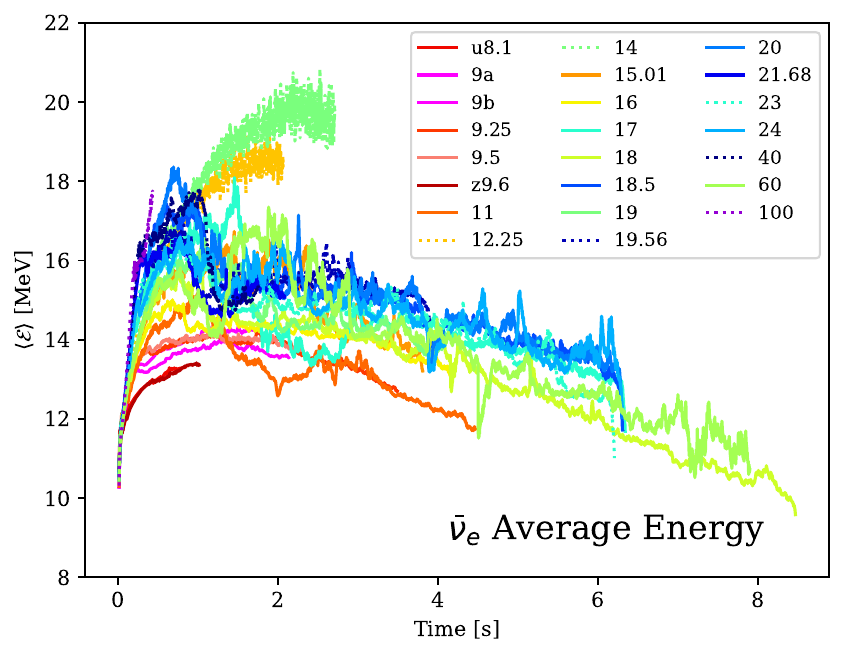}
    \includegraphics[width=0.49\linewidth]{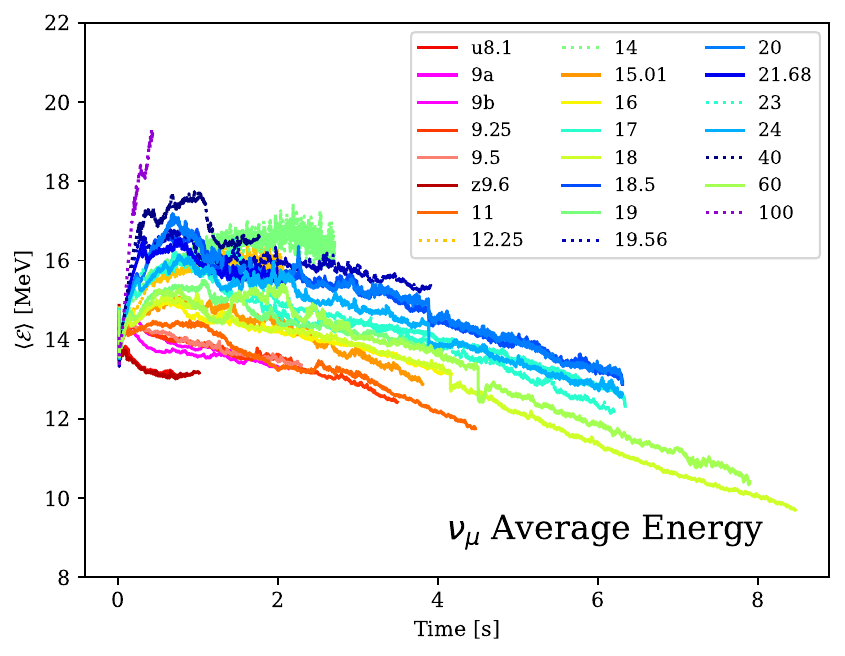}
    \captionsetup{justification=raggedright}
    \caption{Average (mean) energy as a function of time for each model and neutrino species colored compactness. The BH-forming models are plotted with dotted lines. In general, while the $\nu_e$ average energy hovers around a flat mean of 12.5 MeV throughout the simulation, the average energies of the $\bar{\nu}_e$ and $\nu_{\mu}$ species decrease with time, peaking at around 16-18 MeV and decreasing at a rate of around 0.75 MeV per second until the end of the simulation. We also observe an approximate nesting of the average energy as a function of compactness across all species, but due to the higher-amplitude temporal variations in the average energy signal, this trend is not as clear as with the luminosity signals. The nonexploding 12.25 and 14 $M_{\odot}$ models have a clear, distinguishable $\nu_e$ and $\bar{\nu}_e$ average energy evolution that rises faster, oscillates at a higher frequency compared to the exploding models, and reaches greater energies (18-20 MeV) than even the most compact models in this suite of simulations.
    \label{avg_energy}}
    \end{figure*}

In Figure \ref{avg_energy}, we plot the average neutrino energy in MeV as a function of time for each model colored by compactness. Overall, we observe that the average energies seem to fluctuate around a flat mean of around 12.5 MeV for the $\nu_e$ species, whereas the mean of the fluctuations decreases with time at a rate of 0.75 MeV per second for the $\bar{\nu}_e$ and $\nu_{\mu}$ neutrinos. The $\bar{\nu}_e$ and $\nu_{\mu}$ species also have slightly harder spectra across all models with average energies ranging from around 12 to 20 MeV for $\bar{\nu}_e$ neutrinos and 14 to 18 MeV for $\nu_{\mu}$ neutrinos, whereas the average energy for the $\nu_e$ neutrinos falls between 10 and 16 MeV, a result consistent with previous simulations \cite{Mayle1987,Fischer2010,Hudepohl2010,OConnor_2013,Nagakura2020, Nagakura2021, Vartanyan_Burrows2023}. 

In Figure \ref{avg_energy}, we observe that the average energy is roughly monotonic with compactness for all species, but is not as clearly nested as the corresponding luminosity signals, with more overlap in the average energies due to vigorous temporal fluctuations. These temporal variations are less pronounced for the $\nu_{\mu}$ average energy, and lower-compactness progenitors have smoother curves compared to their more compact counterparts. 

Perhaps the most outstanding feature of Figure \ref{avg_energy}, particularly for the $\nu_e$ and $\bar{\nu}_e$ neutrinos, is the average energy evolution of the non-exploding, BH-forming 12.25 and 14 $M_{\odot}$ models that clearly distinguish themselves from the other models. Not only do the average energy curves of these models oscillate quasi-coherently due to the spiral SASI, but they also quickly rise to higher mean energies within the first second and grow to the highest mean energies (around 18 MeV for $\nu_e$ neutrinos and 20 MeV for $\bar{\nu}_e$ neutrinos) of all of the models, even compared to models with higher progenitor compactnesses. These neutrinos obtain higher mean energies due to the significant compression that results from continued accretion onto the PNS that causes the temperatures around the neutrinospheres to rise. Unlike the other models whose average energies either plateau or have a slight decline, those for the 12.25 and 14 models continue to grow until the end of the simulation. Therefore, non-exploding, BH-forming models (i.e., Channel 4 models \cite{Burrows2024_BH}) have a clear and distinguishable average energy fingerprint, enabling us to witness this particular channel of BH formation with neutrino detectors.

\begin{figure*}[htbp]
    \centering
    \includegraphics[width=0.32\linewidth]{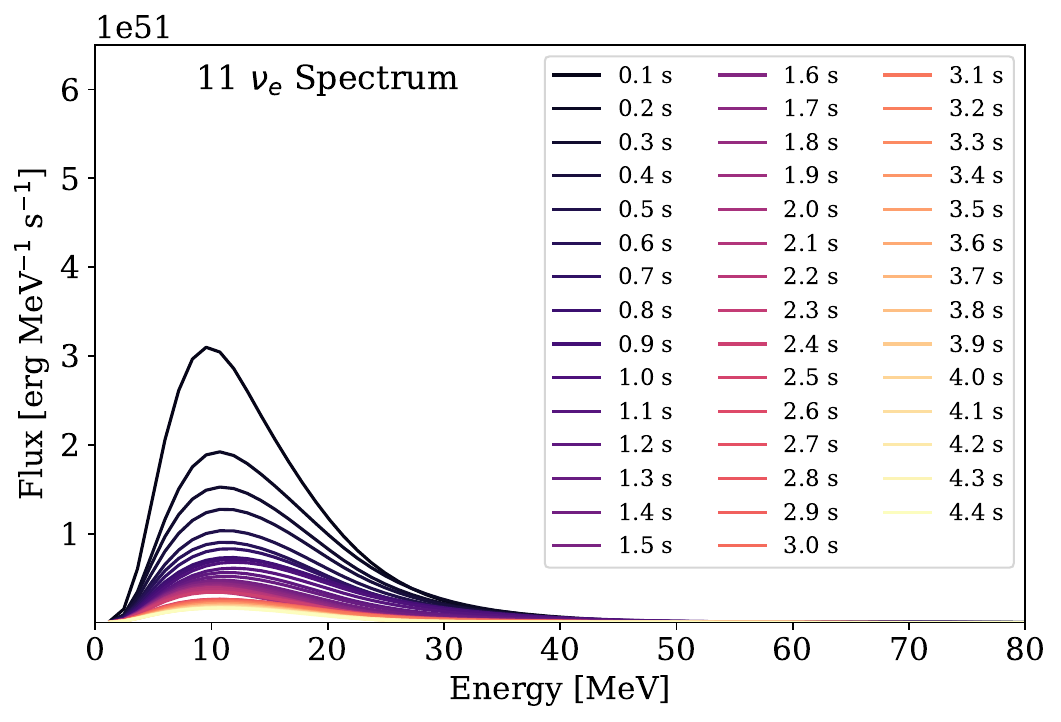}
    \includegraphics[width=0.32\linewidth]{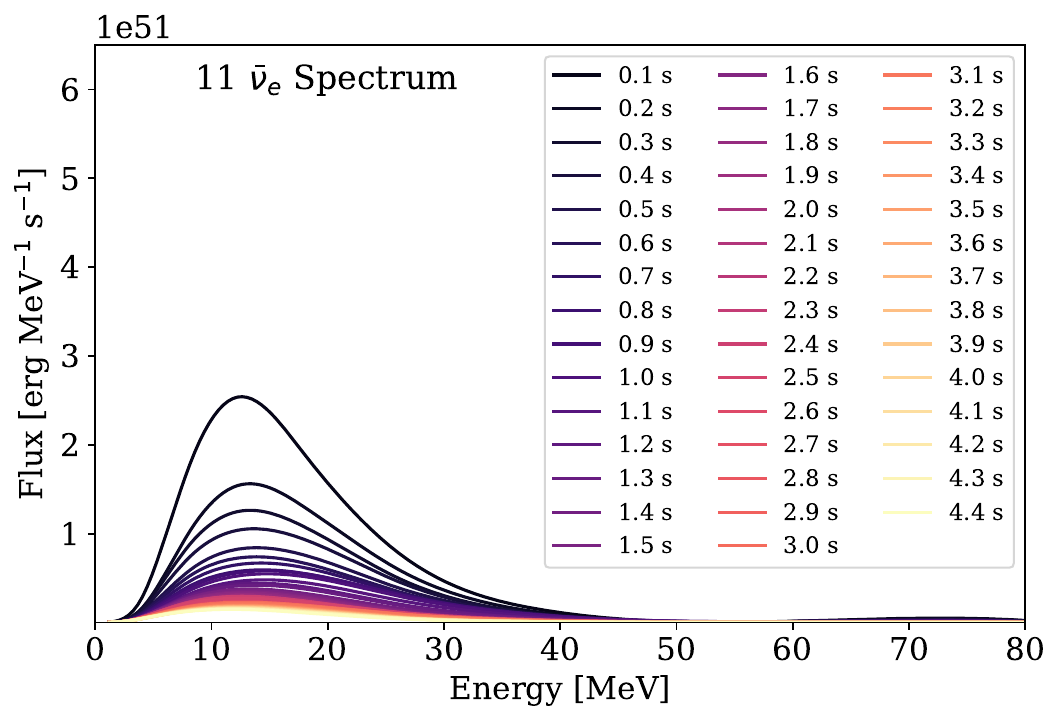}
    \includegraphics[width=0.32\linewidth]{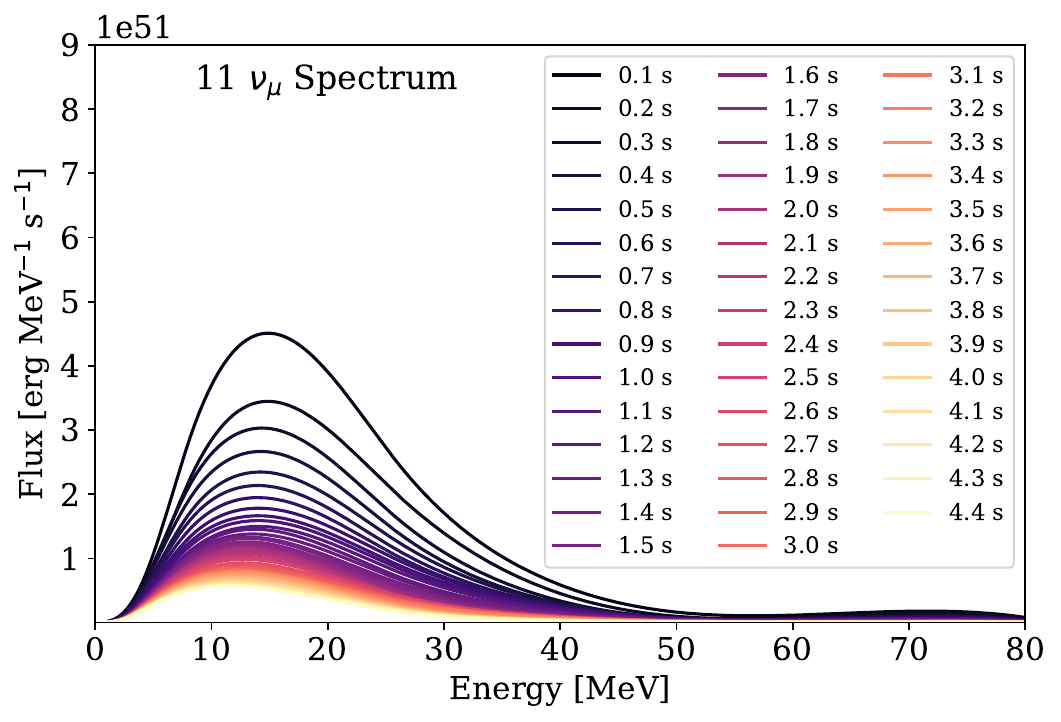}
    \includegraphics[width=0.32\linewidth]{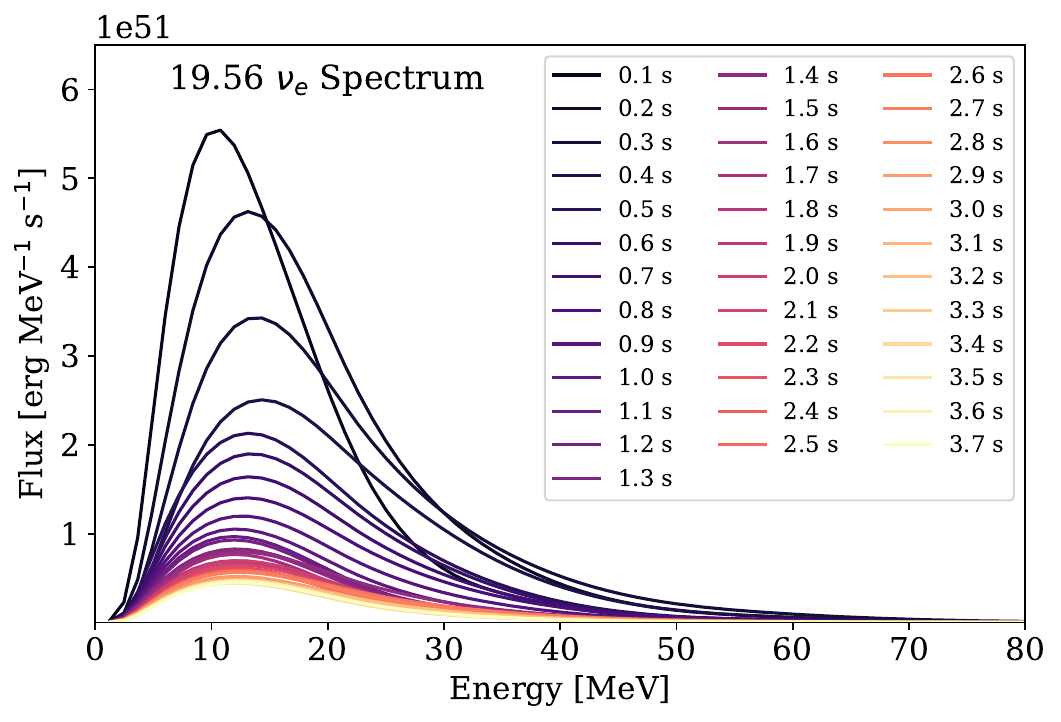}
    \includegraphics[width=0.32\linewidth]{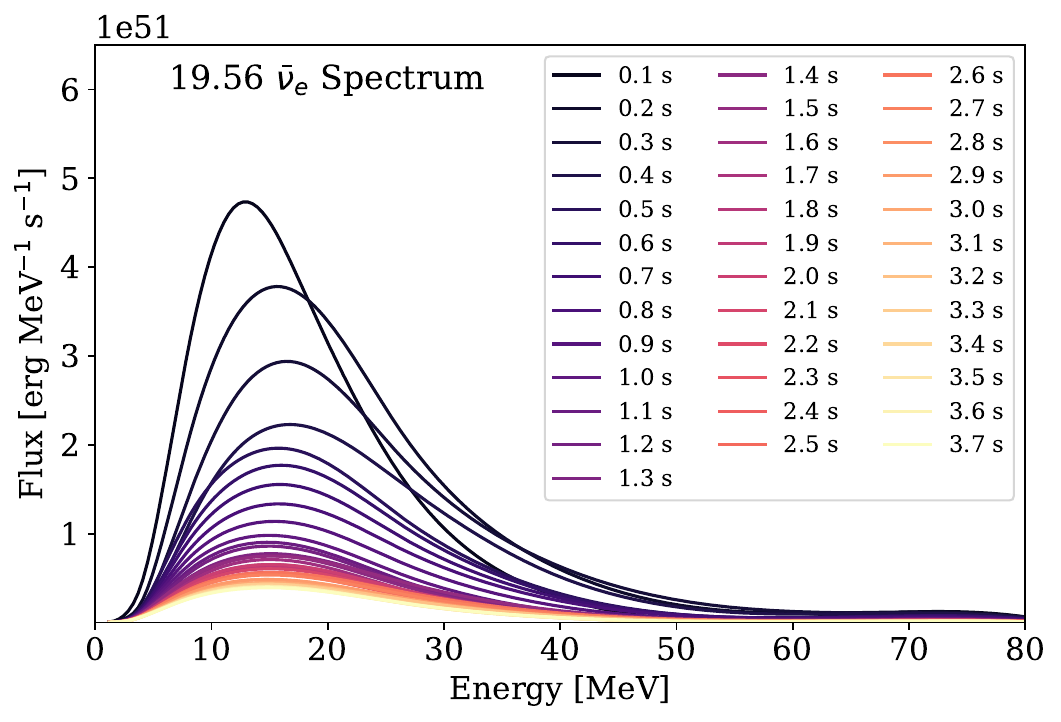}
    \includegraphics[width=0.32\linewidth]{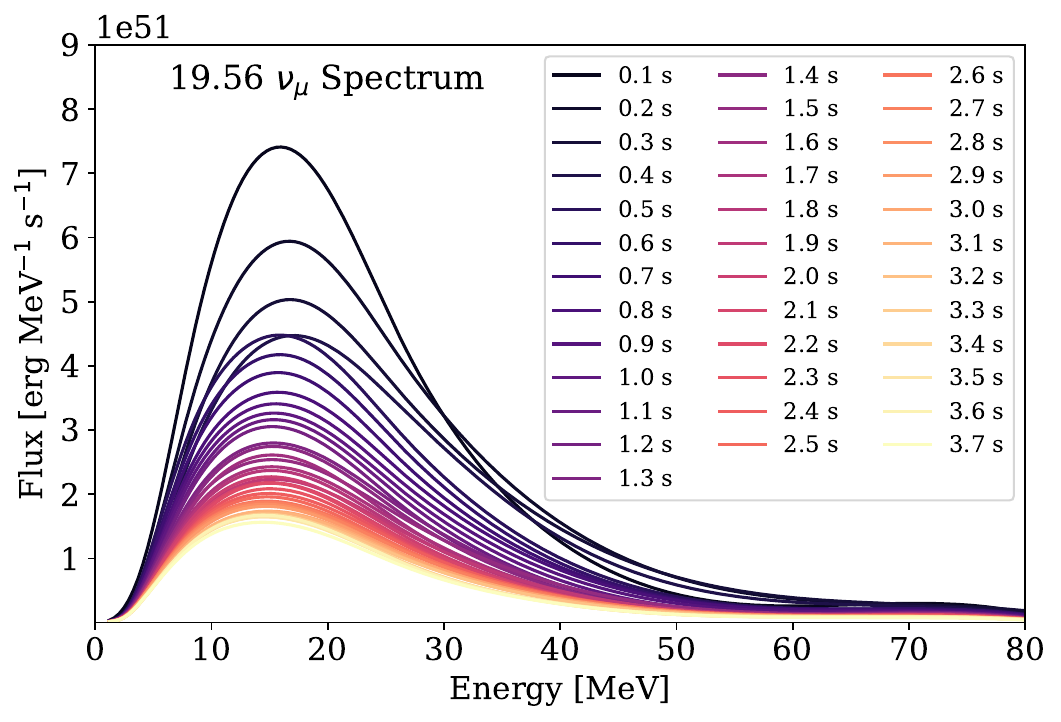}
    \captionsetup{justification=raggedright}
    \caption{Energy spectrum as a function of time for each neutrino species for the 11 and 19.56 $M_{\odot}$ models. The colors indicate the time post-bounce in seconds. Both models' spectra peak around 10-20 MeV across all species. The peak flux is attained at early times, with the flux decreasing with time thereafter. The left-skewed spectra peaked at 10-20 MeV with a higher-energy tail is characteristic of all models in our suite of simulations. \label{spectra}}
    \end{figure*}

Finally, in Figure \ref{spectra}, we plot the neutrino energy spectra for each species for the NS-forming 11 $M_{\odot}$ and BH-forming 19.56 $M_{\odot}$ models as a function of time. Both models' spectra peak around 10-20 MeV for all species, with a higher-energy tail that is more pronounced for the 19.56 $M_{\odot}$ model. For all species, the spectra reach the highest peak at earlier times, and then decrease and shift to slightly lower energy values with time. Though not pictured here, the overall shape of the left-skewed spectra with a peak within 10-20 MeV is characteristic of all models in this suite of simulations, with the main difference between the models' spectra being the height of the peak.

\subsection{Angle-Dependent Luminosity Features}
The luminosity and energy results presented in the preceding sections were derived from angle-averaged neutrino data. In this section, we investigate the angular anisotropy of the neutrino emission, analogously to the exploration of the angular anisotropy in the corresponding GW signals in a previous work \cite{Choi2024}. 

\begin{figure*}
    \centering
    \includegraphics[width=0.48\linewidth]{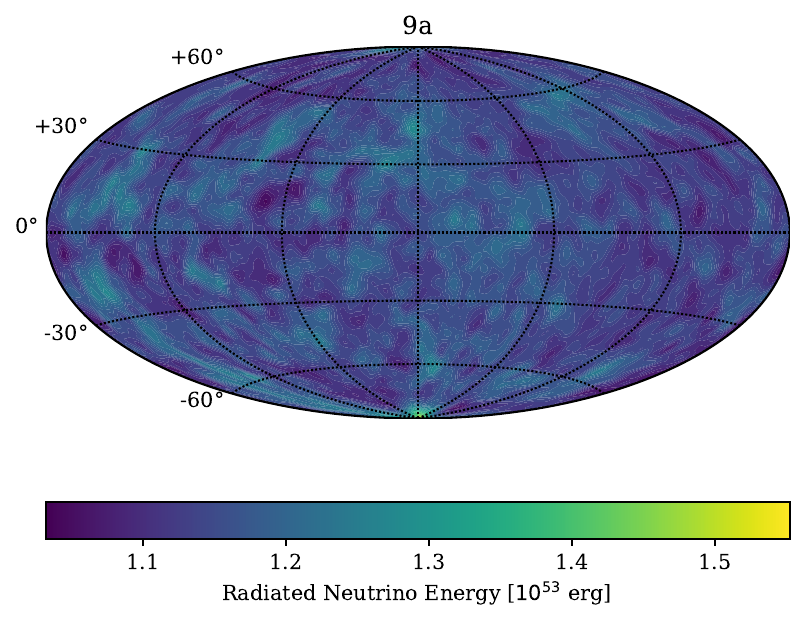}
    \includegraphics[width=0.48\linewidth]{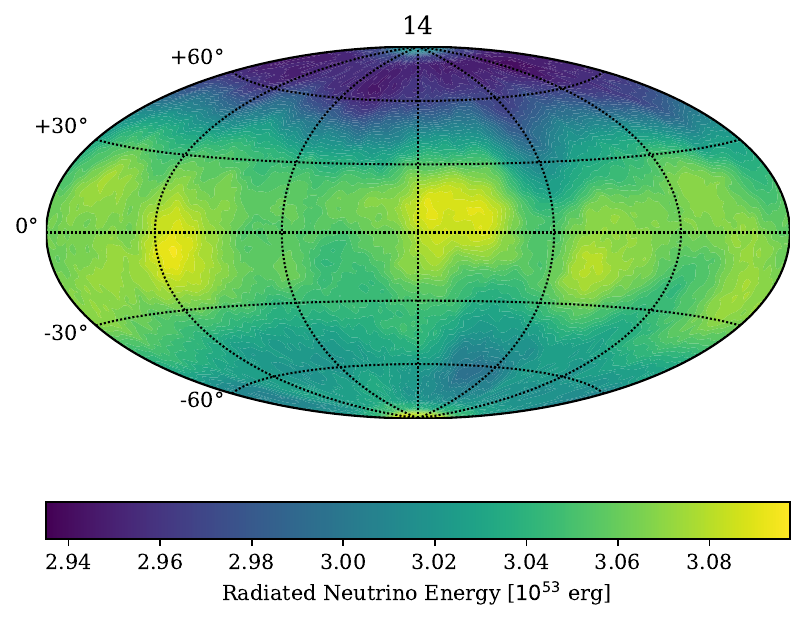}
    \includegraphics[width=0.48\linewidth]{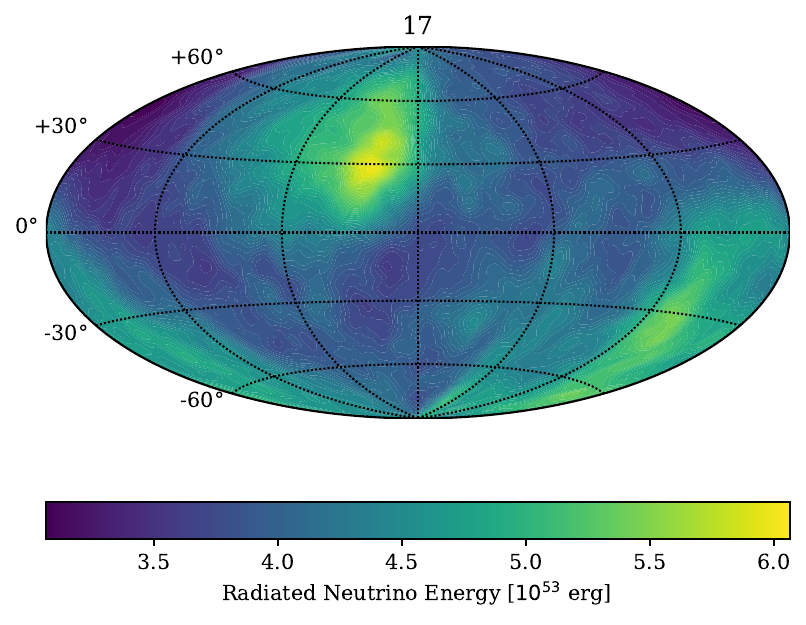}
    \includegraphics[width=0.48\linewidth]{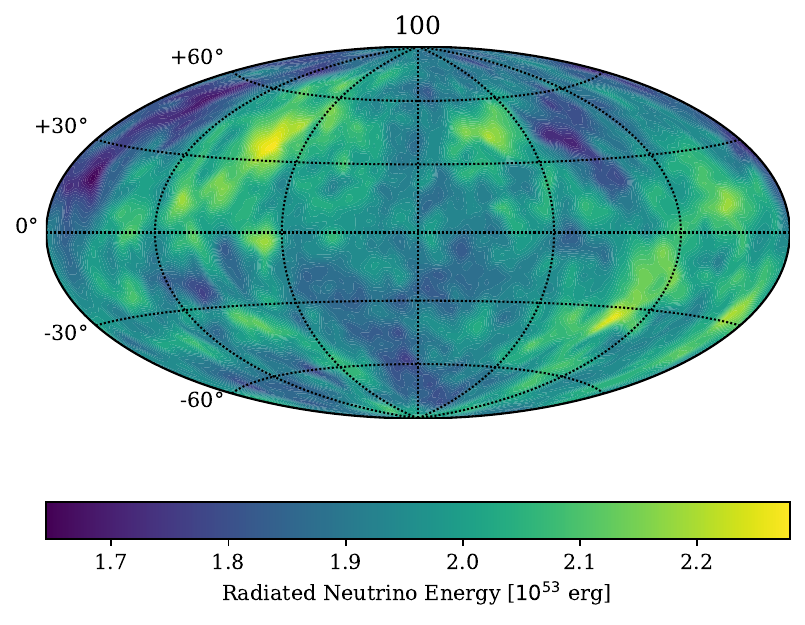}
    \captionsetup{justification=raggedright}
    \caption{Total inferred neutrino energy radiated across all species as a function of viewing angle for the 9a, 14, 17, and 100 $M_{\odot}$ models by the end of the simulations. The neutrino emission for all models is anisotropic, and the qualitative distribution of the anisotropic emission varies between models. In particular, the 17 $M_{\odot}$ model shows a clear hotspot of neutrino emission, whereas the other distributions are more mottled. }
    \label{tone_globes}
\end{figure*}

In Figure \ref{tone_globes}, we plot the total inferred (as if what we see in one direction is emitted isotropically) neutrino energy radiated across all species as a function of viewing angle for the 9a, 14, 17, and 100 $M_{\odot}$ models at the end of the simulation. For each model, there is a significant variation in total neutrino energy inferred as a function of the viewing angle. The qualitative distribution across viewing angles is also different for each model, with some models such as the 9a and 100 $M_{\odot}$ having a more mottled angular distribution and other models, including the very anisotropically-exploding 17 $M_{\odot}$ model \citep{Burrows2024}, displaying a ``hot spot'' in one particular direction. 

\begin{figure*}
    \centering
    \includegraphics[width=0.32\linewidth]{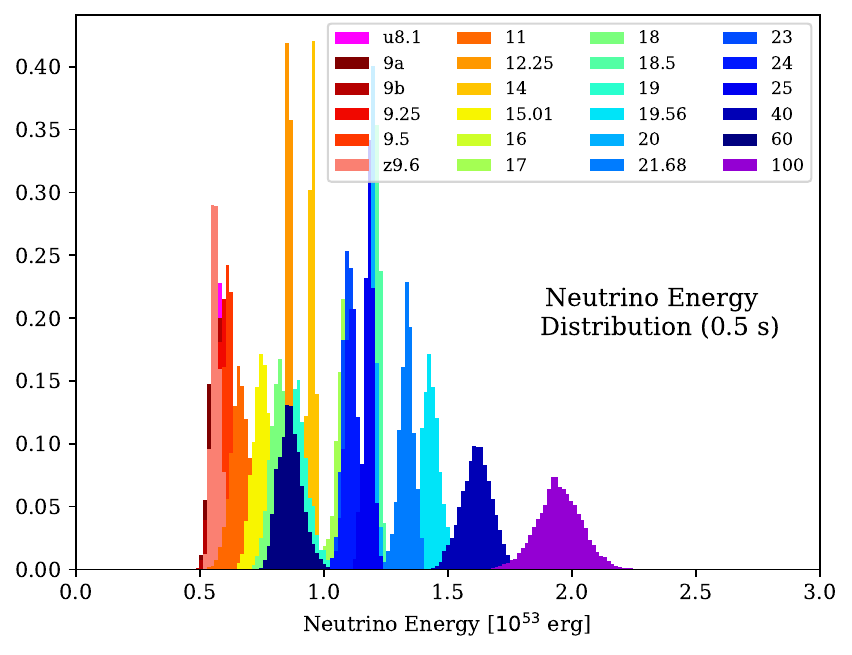}
    \includegraphics[width=0.32\linewidth]{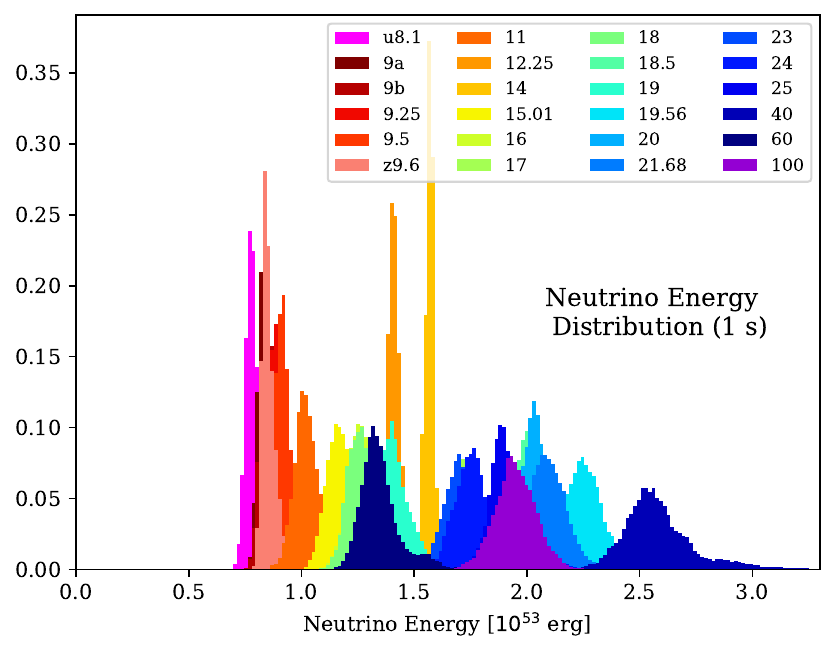}
    \includegraphics[width=0.32\linewidth]{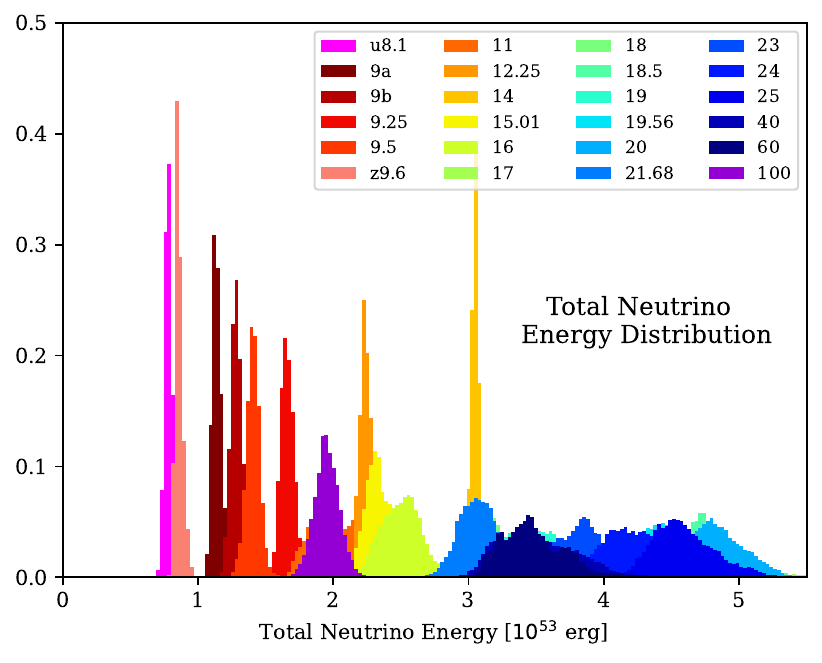}
    \includegraphics[width=0.32\linewidth]{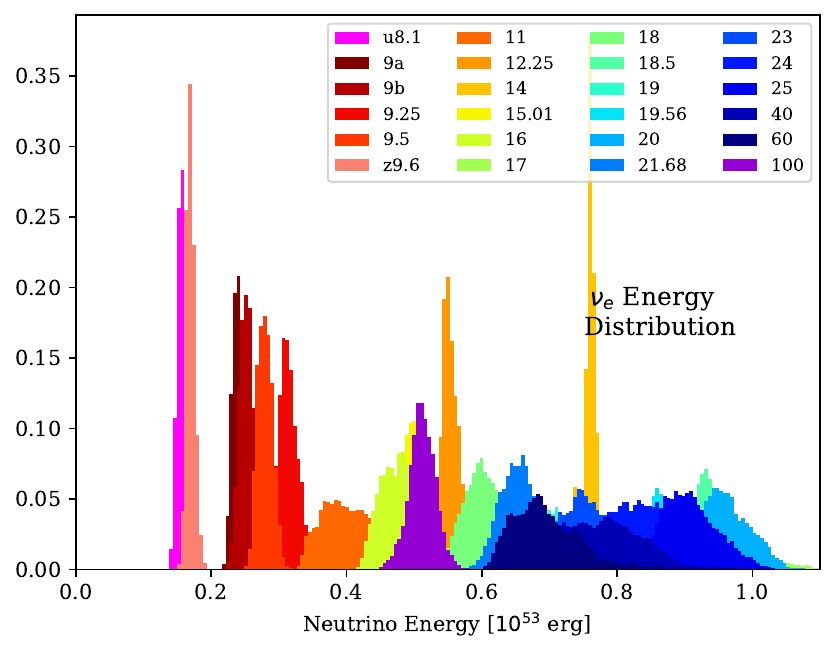}
    \includegraphics[width=0.32\linewidth]{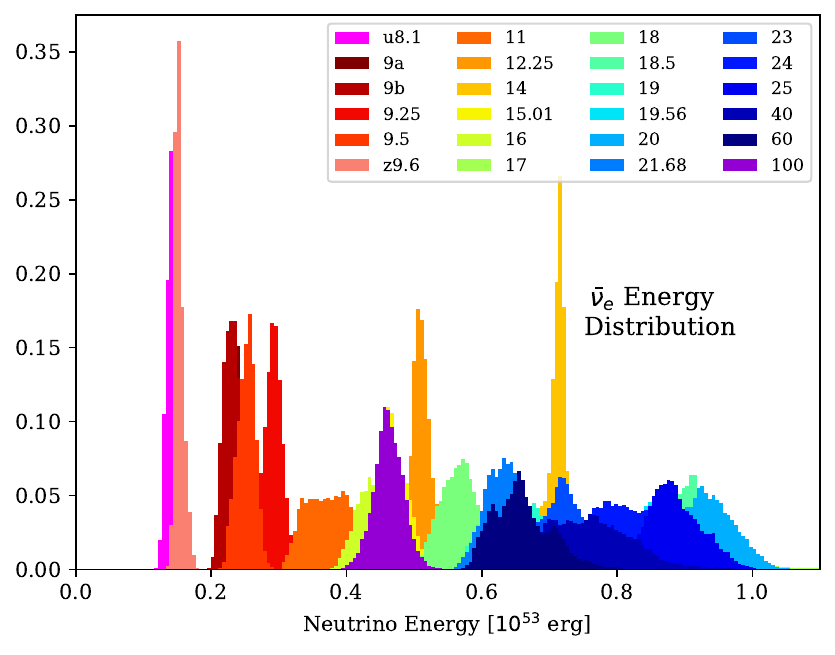}
    \includegraphics[width=0.32\linewidth]{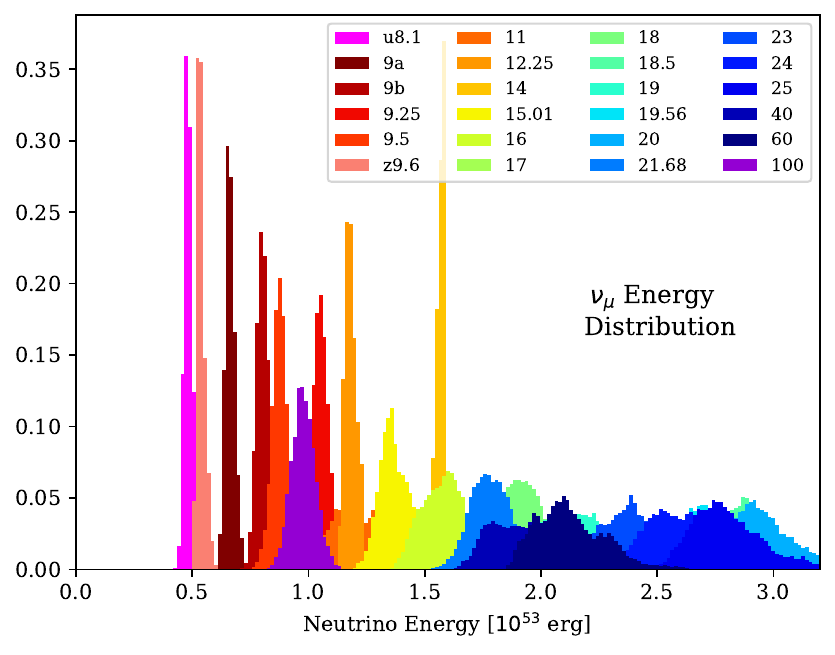}
    \captionsetup{justification=raggedright}
    \caption{Top row: total inferred neutrino energy distribution across all neutrino species and viewing angles for each model after 0.5 seconds, 1 second, and at the end of the simulation. The center of the distribution, as well as its width increases with time, indicating that the energy grows to higher values and becomes more anisotropic with time. Overall, lower-mass models display narrower distributions centered at lower energy values, while higher-mass models have wider distributions at higher energies. The non-exploding 12.25 and 14 models have especially narrow distributions across viewing angles. Bottom row: inferred neutrino energy distribution across all viewing angles for each model for the $\nu_e$, $\bar{\nu}_e$, and $\nu_{\mu}$ neutrinos independently. While the relative distributions in the $\nu_e$ and $\bar{\nu}_e$ energies appear similar between the models, the distributions in the $\nu_{\mu}$ energy are qualitatively more distinct and centered at higher energies, with less similarity in the lower-mass models. However, across all species the non-exploding 12.25 and 14 models have the narrowest distributions, again serving as a distinguishing feature of Chanel 4 BH formation.}
    \label{tone_hists}
\end{figure*}

In order to understand the overall distribution in total inferred emitted neutrino energy, including potential biases associated with particular angles of observation and the evolution of the anisotropy as a function of time, on the top row of Figure \ref{tone_hists}, we decompose these neutrino energy angular distributions into histograms at 0.5 seconds, one second, and by the end of the simulation. These histograms are normalized to one and weighted by $\sin\theta$ where $\theta \in [0, \pi]$ in order to account for the smaller relative contribution to the signal towards the poles. Overall, we observe that the center of the histograms increases in total  energy with time, as well as that the distributions themselves widen with time. This temporal evolution implies that the angular anisotropy increases with time, which also indicates that neutrino emission at earlier times is more isotropic. Overall, higher-mass and more-compact progenitors display wider angular distributions compared to lower-mass and lower-compactness progenitors. Therefore, for lower-mass and non-exploding models, the angle at which we happen to view the supernova may not significantly bias the overall understanding of total neutrino energy radiated and is fairly representative of the neutrino energy in all directions. The wider, overlapping histograms of the higher-mass progenitors may result in more uncertainty in the total neutrino energy radiated of the event when restricted (as we are) to one particular viewing angle. The non-exploding 12.25 and 14 $M_{\odot}$ models, however, display particularly narrow distributions in energy, which again serves as a distinguishing feature of this channel of BH formation. 

On the bottom row of Figure \ref{tone_hists}, we plot a similar histogram distribution for each model, this time for the $\nu_e$, $\bar{\nu}_e$ and $\nu_{\mu}$ species individually by the end of the simulation. Once again, higher-mass progenitors generally have a wider distribution of total neutrino energy emitted as a function of viewing angle, and the non-exploding 12.25 and 14 $M_{\odot}$ models have especially narrow distributions compared to the exploding models. Additionally, the $\nu_e$ and $\bar{\nu}_e$ distributions are qualitatively similar for each model, while the relative distribution in the total radiated $\nu_{\mu}$ energy is different, with less overlap in the distributions of the lower-mass models compared to the other species and peaking at higher energies. 

\section{Detector Signals}
\subsection{Detector Summary}

While there are many neutrino detectors in development including HALO-2 \cite{Duba2008}, XENONnT \cite{Xenon2024, Kopec2022}, DARWIN \cite{Aalbers2016}, DarkSide-20 kt \cite{Darkside2021}, RES-NOVA \cite{Alloni2025}, and COSINUS \cite{angloher2024}, for the present study we focus on the detectors Super-Kamiokande (SK) \cite{Abe_2016}, JUNO \cite{An_2016}, DUNE \cite{Ankowski2016}, and IceCube \cite{Abbasi2011, Kopke_2011}. We provide a brief overview of each of these detectors, and then discuss a few of the observable features of our models with these detectors. 

SK is a 32.5 kton water Cherenkov detector located in Japan. The primary detection channel of SK is via the inverse beta decay (IBD) reaction:
 \begin{equation}
    \label{IBD}
        \bar{\nu}_e + p \rightarrow e^+ + n\, ,
    \end{equation}
making it most sensitive to $\bar{\nu}_e$ neutrinos. Its successor Hyper-K is expected to have a fiducial volume and, as a consequence, an event rate around eight times greater than SK. Similarly, the 20-kton liquid scintillator JUNO in China is also most sensitive to $\bar{\nu}_e$ neutrinos via the IBD reaction.

DUNE is a 40 kton liquid argon detector that is most sensitive to $\nu_e$ neutrinos via the charged current reaction with argon:
\begin{equation}
    \label{nu_ar}
        \nu_e + ^{40}Ar \rightarrow e^- + ^{40}K^*\, . 
\end{equation}

Lastly, IceCube is a 3.5 Mton detector located deep in pure ice in Antarctica. As with SK and JUNO, it would be most sensitive to $\bar{\nu}_e$ neutrinos via the IBD reaction, but employs solid ice rather than water. 

A unique property of neutrinos is their ability to oscillate between different flavors during propagation. As a consequence, the relative mix of neutrino species emitted at the supernova may not match the relative mix of neutrinos detected at Earth. While the exact neutrino oscillation model remains an active area of research, in this paper, we present the results of the event rates in each detector for three distinct cases: assuming no oscillation model, assuming adiabatic flavor conversion due to Mikheyev-Smirnov-Wolfenstein (MSW) effects for the normal mass hierarchy, and assuming adiabatic MSW oscillation effects for the inverted mass hierarchy. For the survival probability $p$ of neutrinos and $\bar{p}$ of anti-neutrinos, the neutrino and anti-neutrino flux at the Earth (given by $F_i$ and $\bar{F}_i$, respectively) are calculated as \citep{Dighe2000}:
\begin{equation}
\begin{aligned}
\label{MSW_flux}
    F_e &= pF_e^0 + (1-p)F_{\mu}^0 \\
    \bar{F}_e &= \bar{p}\bar{F}_e^0 + (1-\bar{p})\bar{F}_{\mu}^0 \\
    F_{\mu} &= \frac{1}{2}(1-p)F_e^0 + \frac{1}{2}(1+p)F_{\mu}^0 \\
    \bar{F}_{\mu} &= \frac{1}{2}(1-\bar{p})\bar{F}_e^0 + \frac{1}{2}(1+\bar{p})\bar{F}_{\mu}^0 
    \end{aligned}
\end{equation}
where $F^0$ and $\bar{F}^0$ denote the flux of the neutrinos and anti-neutrinos respectively without flavor conversion. For the normal mass hierarchy, the neutrino and anti-neutrino survival probabilities can be written as
\begin{equation}
\begin{aligned}
\label{p_NMO}
    p &= \sin^2\theta_{13} \\
    \bar{p} &= \cos^2\theta_{12}\cos^2\theta_{13} \, . 
    \end{aligned}
\end{equation}
The probabilities for the inverted mass hierarchy are
\begin{equation}
\begin{aligned}
\label{p_IMO}
    p &= \sin^2\theta_{12}\cos^2\theta_{13} \\
    \bar{p} &= \sin^2\theta_{13}  
    \end{aligned}
\end{equation}
where the values of mixing parameters $\theta_{12}$ and $\theta_{13}$ are set by $\sin^2\theta_{12} = 2.97 \times 10^{-1}$ and $\sin^2\theta_{13} = 2.5 \times 10^{-2}$ \cite{Capozzi2017}. We implement these flavor transformations using the SNEWPY software \citep{snewpy}, and provide details of this implementation in the Appendix. 

\subsection{Neutrino Detector Results}
In this section, we provide an example set of calculations for the neutrino event rates from our models as observed by the JUNO, DUNE, SK, and IceCube detectors using the SNEWPY software \citep{snewpy}. The purpose of these calculations is to determine whether the trends identified in the luminosity curves in Section III are manifest in neutrino detector signals. We emphasize that this discussion is an overview of detection capabilities and is only meant to serve as a starting point for a more in-depth discussion of neutrino signal detectability reserved for future work. 

\begin{figure*}[htbp!]
    \centering
    \includegraphics[width=0.32\linewidth]{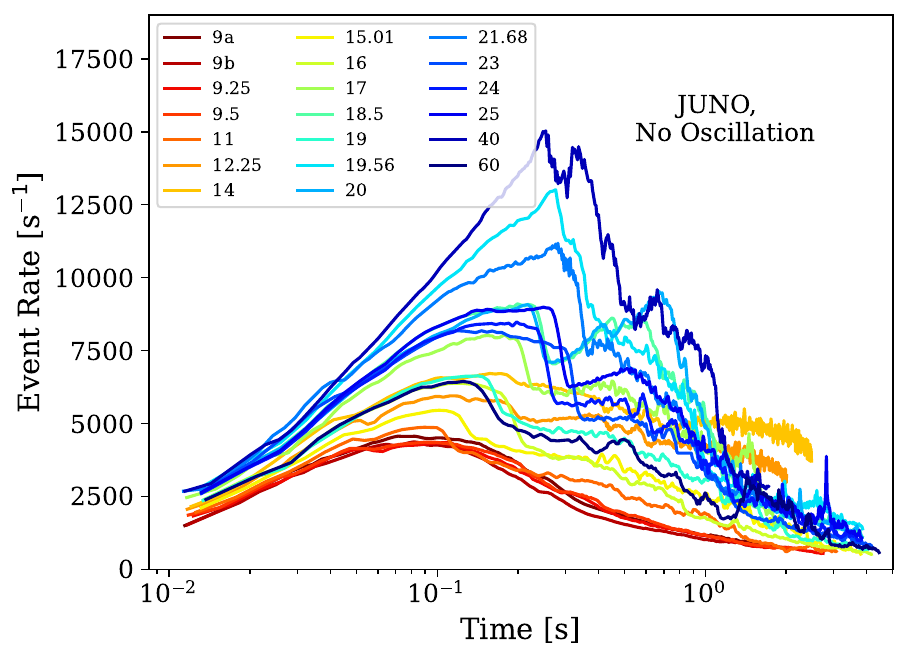}
    \includegraphics[width=0.32\linewidth]{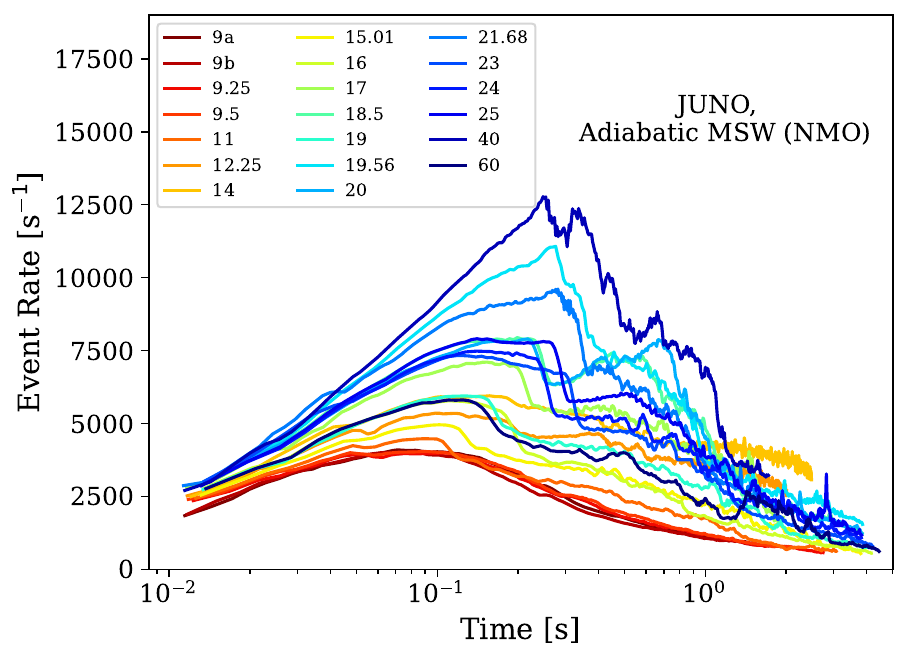}
    \includegraphics[width=0.32\linewidth]{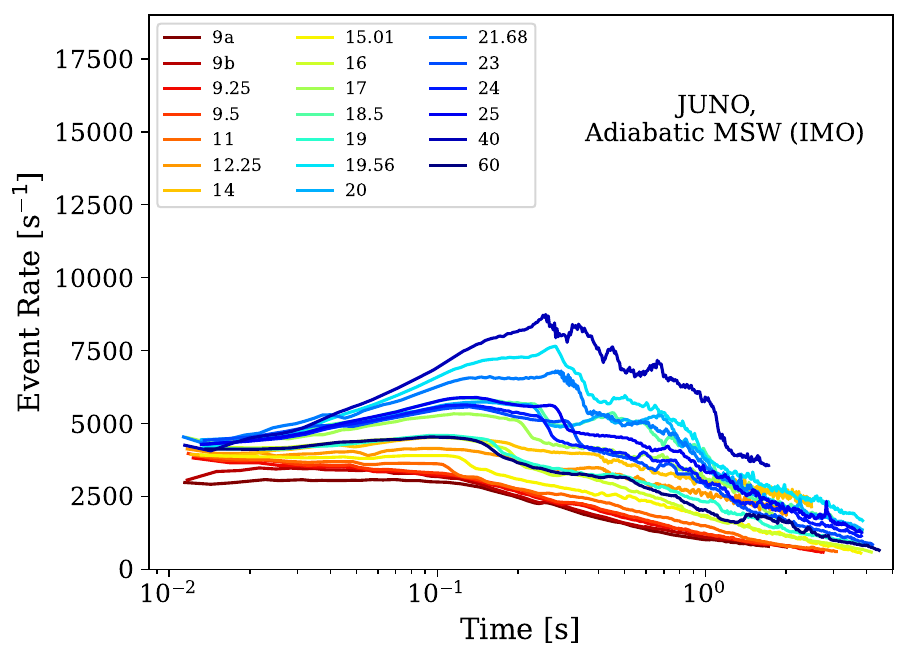}
    \includegraphics[width=0.32\linewidth]{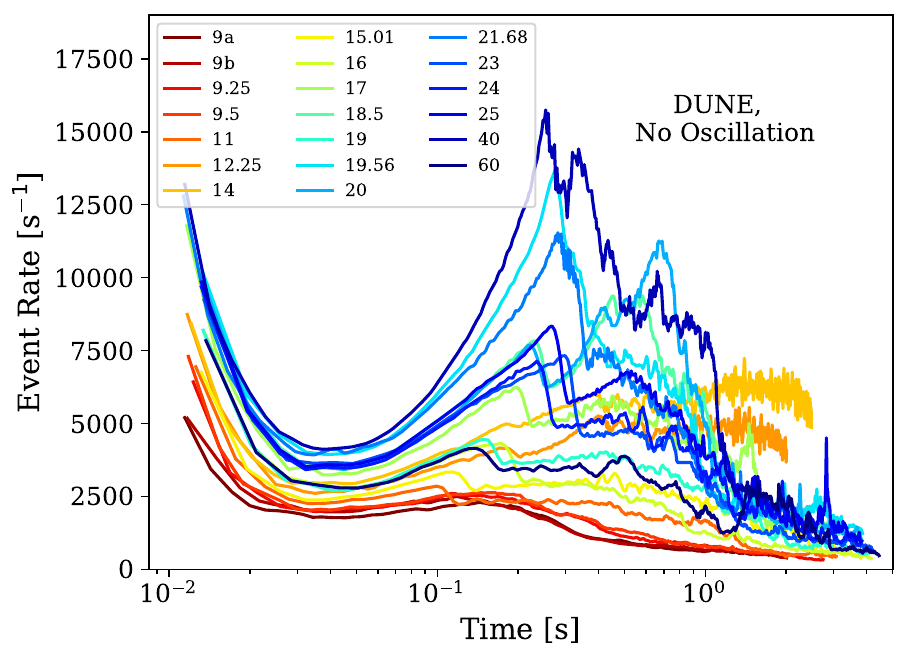}
    \includegraphics[width=0.32\linewidth]{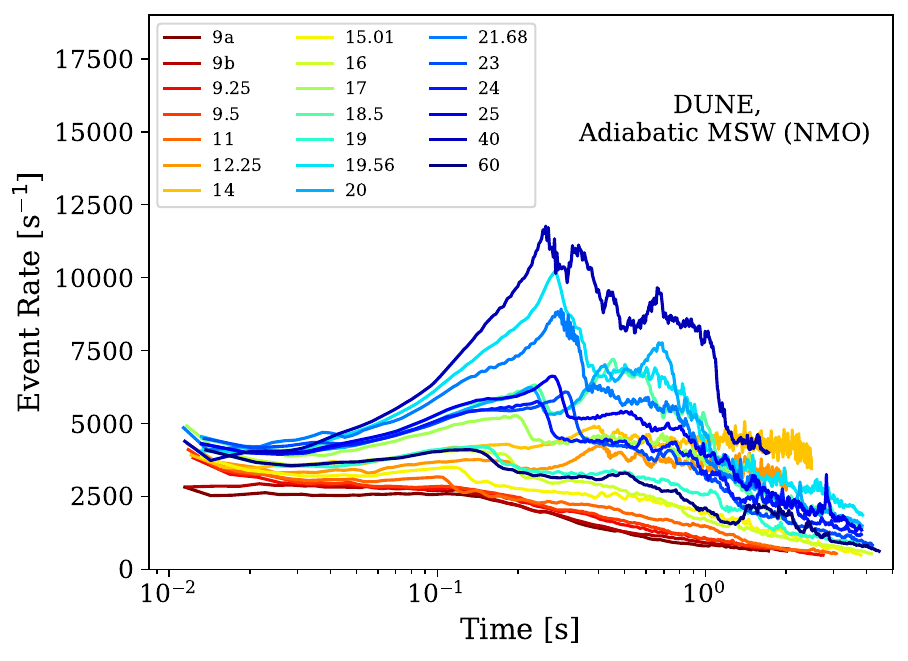}
    \includegraphics[width=0.32\linewidth]{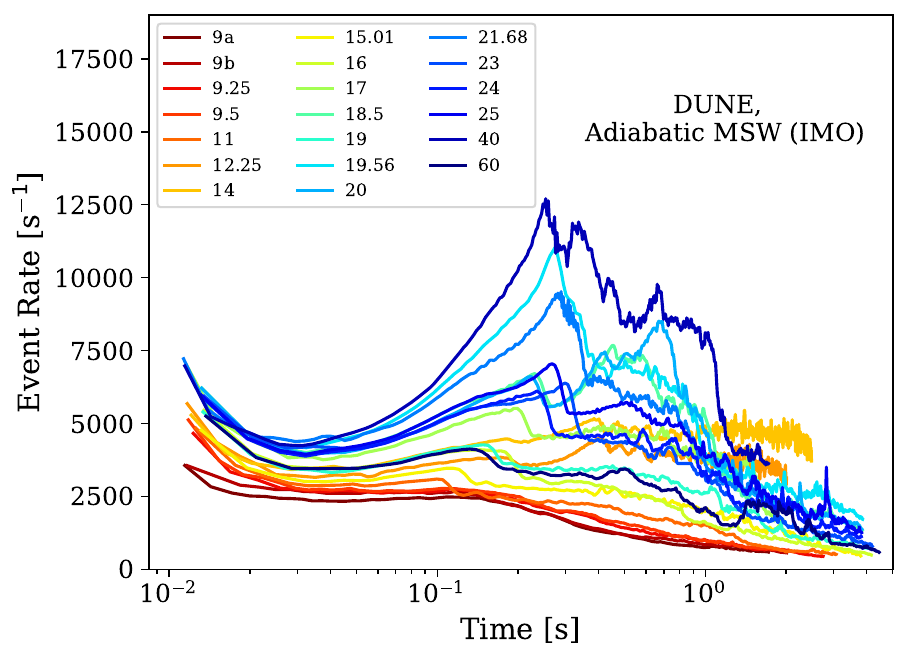}
    \includegraphics[width=0.32\linewidth]{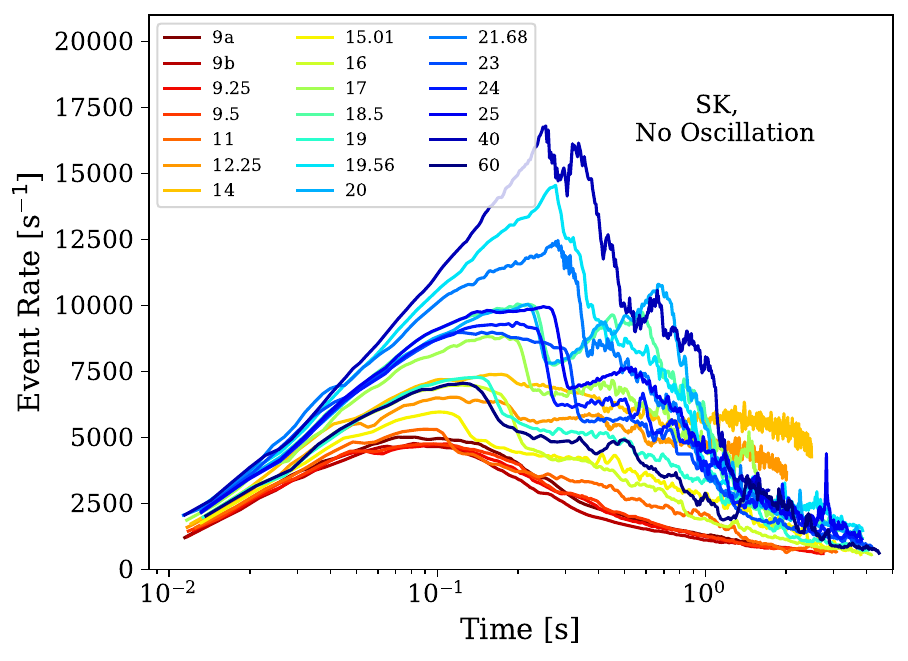}
    \includegraphics[width=0.32\linewidth]{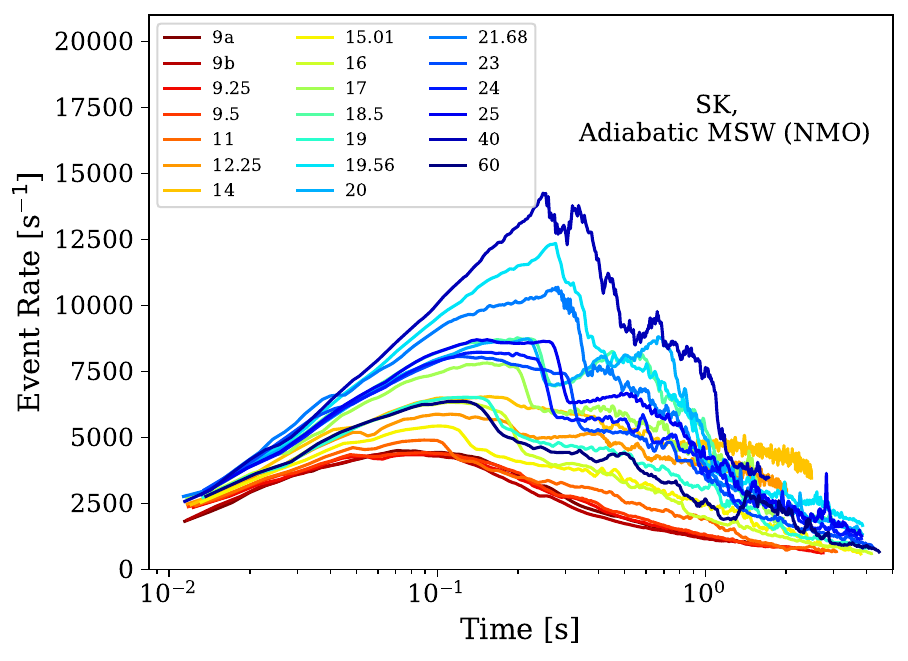}
    \includegraphics[width=0.32\linewidth]{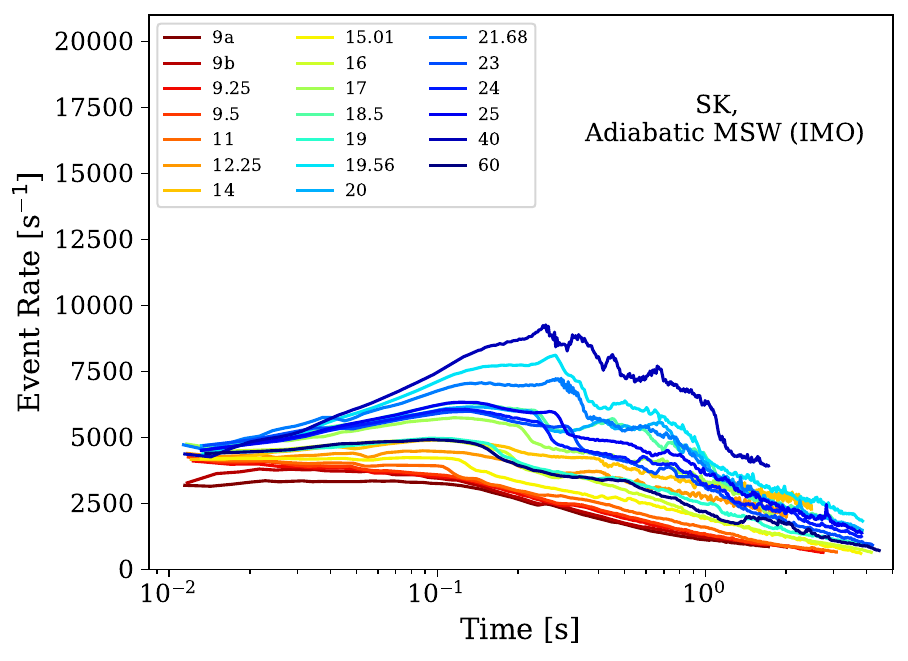}
    \includegraphics[width=0.32\linewidth]{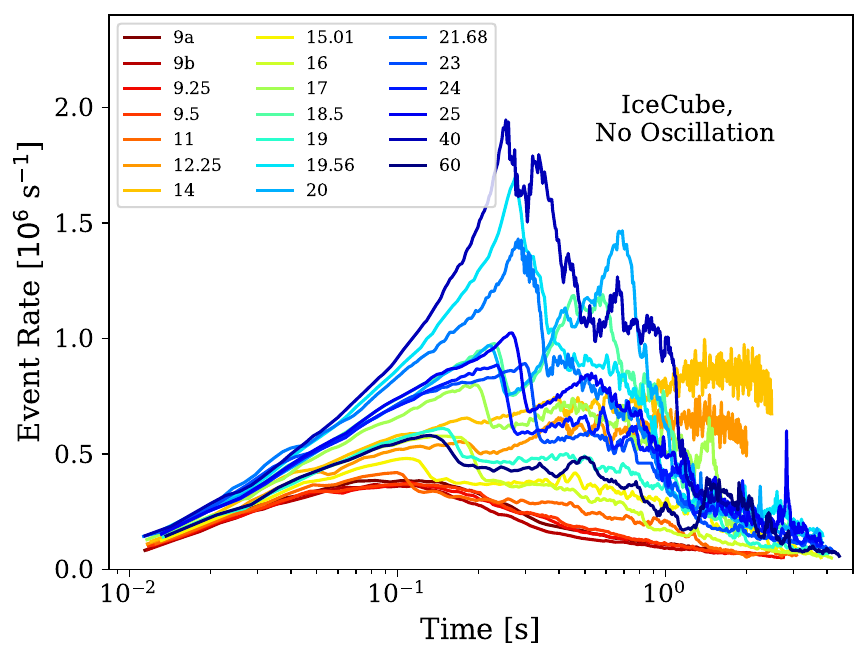}
    \includegraphics[width=0.32\linewidth]{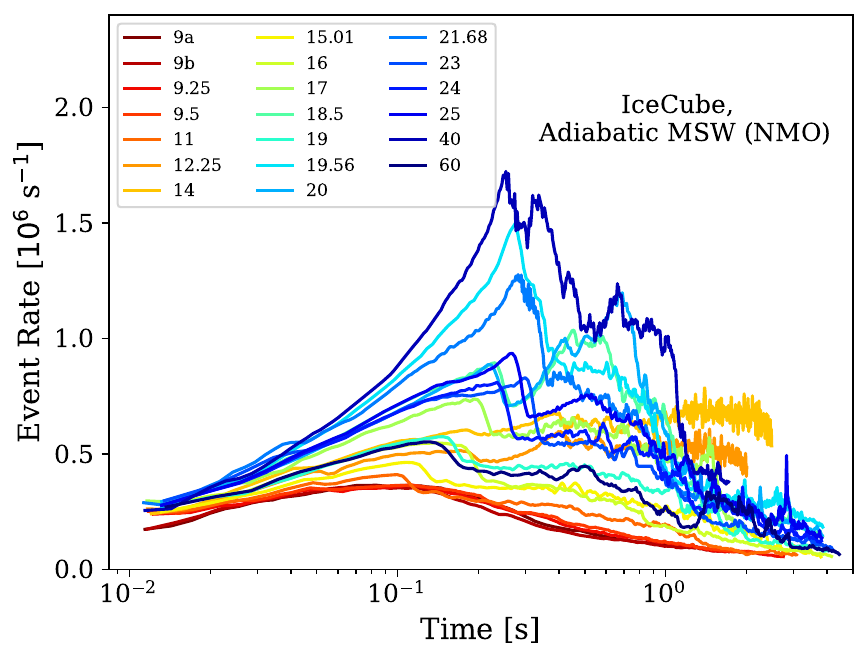}
    \includegraphics[width=0.32\linewidth]{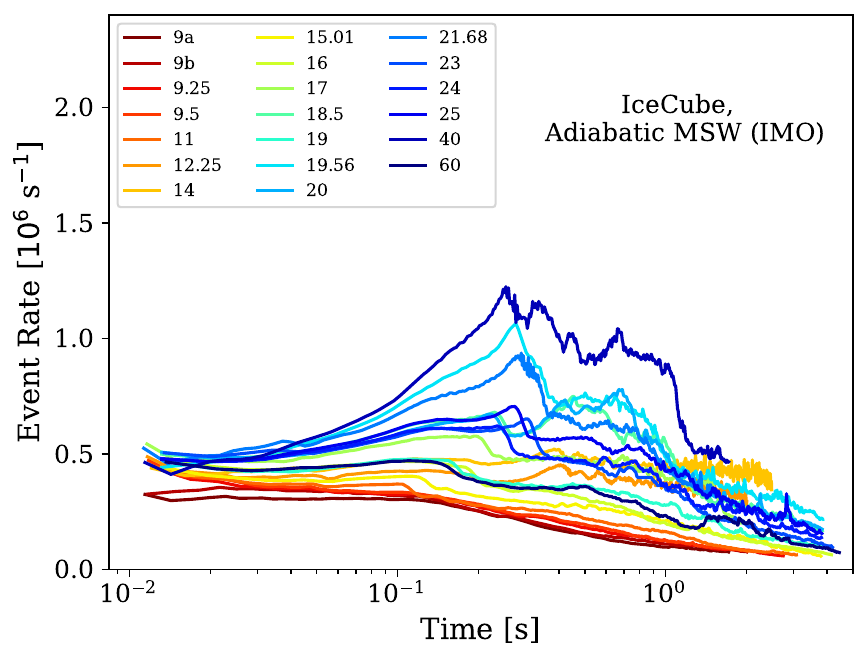}
    \captionsetup{justification=raggedright}
\caption{Neutrino lightcurves for the main detection channels in the JUNO, DUNE, SK, and IceCube detectors assuming no oscillation (left column), adiabatic MSW oscillation with normal mass ordering (NMO, middle column), and adiabatic MSW with inverted mass ordering (IMO, right column), for 20 of our models at a distance of 10 kiloparsecs. IceCube records the highest event rates among these detectors, but the other detectors still experience events on the order of many thousands. Overall, all detectors are able to observe the accretion-powered luminosity peak and plateau as well as the decay from the plateau due to cooling. As the detector most sensitive to the $\nu_e$ species, DUNE is the only detector that shows evidence of the decay from the breakout burst, but this breakout signature is difficult to resolve for the normal mass ordering MSW oscillation. The progenitor-dependent nesting is still maintained each of the detectors' signals, enabling us to infer progenitor compactness directly from neutrino observations. Overall, event rates decrease between normal-mass ordering oscillation and no oscillation, and decrease even further for inverted mass hierarchy MSW oscillation for the IBD-channel detectors. Additionally, for the IBD-channel detectors, the spiral SASI of the non-exploding 12.25 and 14 models is evident assuming no neutrino oscillations, but the spiral SASI is buried under the other models' lightcurves for each neutrino oscillation model. 
\label{detector_plots}}
    \end{figure*}

In Figure \ref{detector_plots}, we plot the neutrino event rates for each detector and 20 of our models under no oscillation model (left column), normal-ordered adiabatic MSW oscillation (middle column), and inverted mass order MSW oscillation (right column), assuming a distance of 10 kiloparsecs. Across all progenitors and oscillation models, IceCube has the highest event rate, on the order of $10^6$ events per second, whereas the other detectors experience events on the order of $5 \times 10^3$ to around $1.5\times10^4$ events per second. In addition to detecting the neutrino signal, from Figure \ref{detector_plots} we can also see that all detectors are able to follow several stages of the CCSN process, including the accretion-powered luminosity rise and the decay due to cooling. The decay from the initial breakout burst is detectable with DUNE due to its unique sensitivity to $\nu_e$ neutrinos, but only under certain oscillation models. The compactness-dependent nesting of the luminosity curves observed in Figure \ref{luminosities24} is also present across all detectors and oscillation models, with the most compact progenitors reaching event rates two to three times greater than the event rates of the least-compact progenitors across all detectors and oscillation models. This nesting, especially within the detector signal, enables us to approximately distinguish progenitor core structure based on event rate. We can also observe the clear Channel 4 BH-forming spiral SASI signature with all detectors, though this signature is buried with the inverted hierarchy oscillation model. 

As mentioned previously, neutrino oscillation remains an active area of research, and a galactic CCSN event has the potential to constrain the neutrino oscillation model. There have been many studies in the literature investigating neutrino oscillation constraints from CCSN events \cite{Dighe2000, Mirizzi2016}. By looking at Figure \ref{detector_plots}, we can see the oscillation model has a clear impact on the event rate across all models and detectors, with the inverted hierarchy MSW model resulting in the lowest event rates for the IBD-channel detectors. With electromagnetic and other observational constraints on the progenitor mass, the discrepancy between the expected event rate with no oscillation and the observed event rate may enable us to constrain the oscillation model. Another potential signature of the oscillation model can be seen particularly with DUNE. As seen in the second row of Figure \ref{detector_plots}, the decay from the breakout burst appears for the plots with inverted hierarchy oscillation and no oscillation, but is absent for the normal hierarchy oscillation model. Therefore, whether or not the breakout burst is observed with DUNE may also constrain the oscillation model, a result consistent with previous studies on CCSN neutrino detectability \cite{OConnor_2013,Ankowski2016,Wallace_2016, seadrow2018}. 

Finally, we note that the event rates in Figure \ref{detector_plots} are plotted assuming no detector noise. In reality, most of these detectors are limited by Poisson noise (with IceCube limited by background). For detectors limited by Poisson noise, the signal-to-noise ratio is determined straightforwardly by signal bin width. Wider bins result in higher signal-to-noise ratios, but with the tradeoff of lower temporal resolution. Therefore, in practice, the detectors will be able to follow the overall shapes of those presented in Figure \ref{detector_plots} and, thus, still can be used to infer processes such as accretion and cooling, but the exact time-dependent fluctuations will be uncertain and difficult to distinguish from noise unless the event occurs close by. Additionally, given the varying sensitivities of each detector to different neutrino species, combining the signals from multiple detectors will provide a more holistic picture of the supernova lightcurve from Earth \cite{GalloRosso2021}, a project we reserve for future work. 

\section{Correlation with Gravitational-Wave Signal}
In this section, we highlight correlations between the neutrino and gravitational-wave (GW) signals across the 24 models, and discuss their implications for CCSN behavior as well as the NS equation of state. A more in-depth discussion of the GW signals from this suite of simulations can be found in previous work \cite{Choi2024}. Overall, the onset of the neutrino emission is expected to occur at the same time as the onset of the GW signal, thus enabling the use of the timing of the neutrino signal to locate and distinguish the GW signal from a noisy background \cite{Nakamura2016, Shibagaki2021}. 

\begin{figure*}[htbp!]
    \centering
    \includegraphics[width=0.49\linewidth]{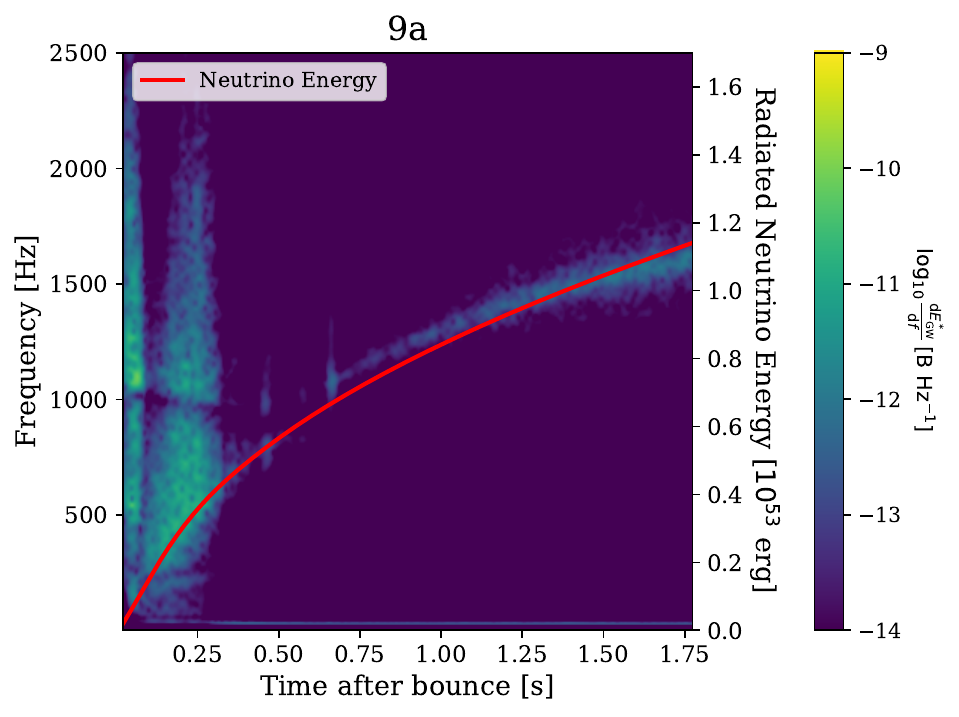}
    \includegraphics[width=0.49\linewidth]{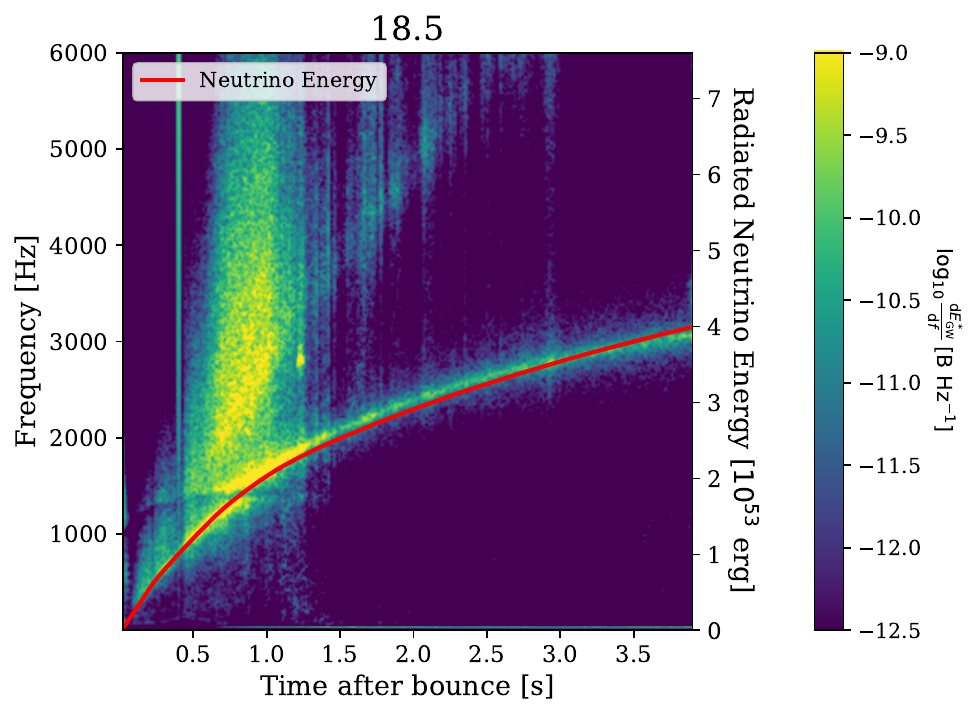}
    \includegraphics[width=0.49\linewidth]{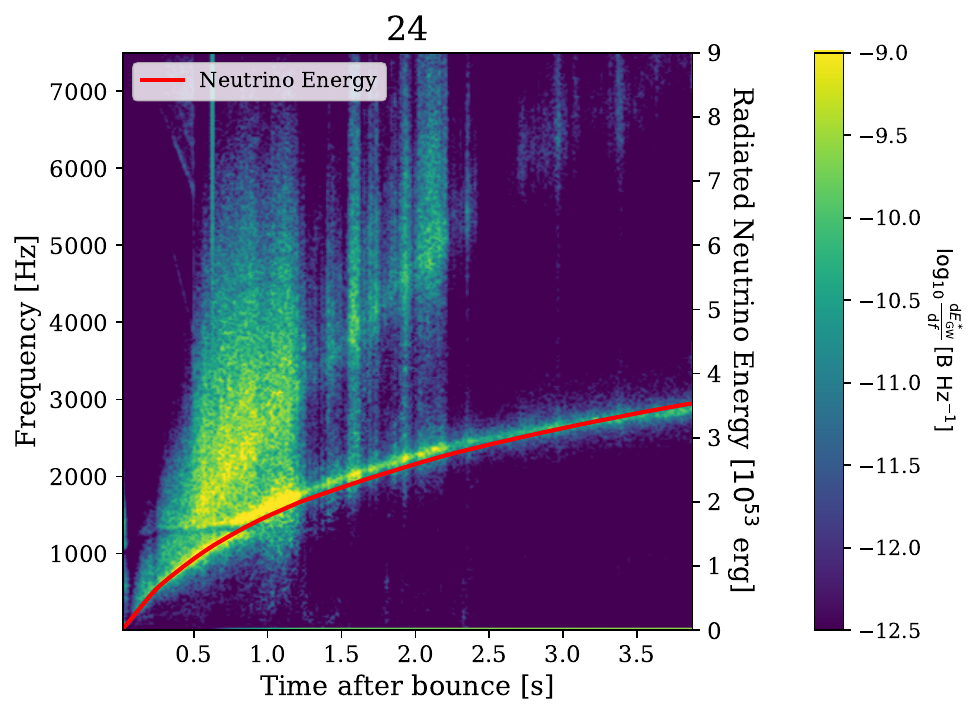}
    \includegraphics[width=0.49\linewidth]{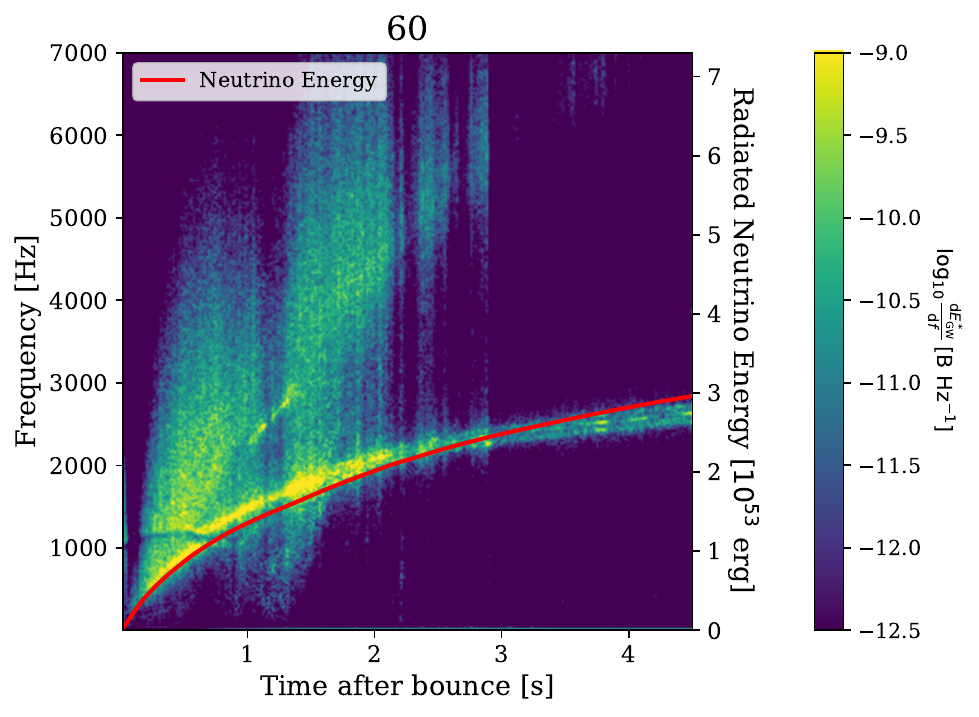}
    \captionsetup{justification=raggedright}
    \caption{Frequency content versus time spectrogram of the GW signal for the 9a, 18.5, 24 and 60 models. Plotted in red is the angle-averaged radiated neutrino energy from Figure \ref{tot_energy} for each model. The rise in the GW f-mode is almost exactly correlated with the total neutrino energy radiated. This relationship occurs because the shrinking core due to neutrino loss ``rings'' at a higher f-mode frequency. 
    \label{tone_fmode}}
    \end{figure*}

In Figure \ref{tone_fmode}, we plot the frequency content versus time GW spectrograms for the 9a, 18.5, 24, and 60 models, along with the total neutrino energy from Figure \ref{tot_energy}. The narrow, high-power band in the spectrogram corresponds to the evolution of the GW f-mode, generated by the ``ringing'' of the PNS as accretion plumes strike its surface. As demonstrated with the comparison to the total neutrino energy plotted in red, the frequency of the GW f-mode evolves in time almost identically to the temporal evolution of the total cumulative neutrino emission. This clear correlation enables us to witness the shrinking of the PNS due to neutrino emission, thus causing the GW f-mode to ring at higher frequencies when bombarded by plumes of accretion. 

Previous CCSN GW studies have revealed that the evolution of the frequency GW f-mode follows different trajectories depending on the underlying equation-of-state of the PNS \cite{Morozova2018}. Therefore, given the correlation between the GW f-mode evolution and the radiated neutrino energy, we expect the neutrino energy evolution to also be dependent on the equation-of-state. While the relative weakness of the CCSN GW signal may make the evolution of the GW f-mode difficult to detect, the correlation in the temporal evolution of both the radiated neutrino energy and the f-mode frequency may enable the use of radiated neutrino energy as a more detectable indicator of the equation-of-state.

\section{Conclusions}
In this paper, we analyzed and highlighted the key features of the neutrino emission from the largest and longest-running suite of 3D CCSN simulations which can serve as scientific targets for the neutrino detector community.

When examining the angle-averaged luminosities for each progenitor model and neutrino species, we explored the various phases of neutrino emission predicted by modern core-collapse supernova theory: the breakout burst, the decay from the breakout burst, the accretion phase to peak and decay, and Kelvin-Helmholtz cooling. The formation of a BH is observable from the abrupt end to the luminosity signal across all species, and the particular case of non-exploding BH-formers are easily distinguished by their oscillatory behavior at late times (after one second) due to the spiral SASI. The magnitude of the luminosity across all species is monotonic with compactness, enabling us to constrain properties of the progenitor core structure from the luminosity signal. Similarly, while the height of the breakout burst peak is independent of progenitor properties, the width increases with progenitor compactness. Due to the monotonic nesting of the luminosity signal with compactness, we also found monotonic correlations between the luminosity peak height across all species as well as a correlation between the total energy radiated in neutrinos with the final NS gravitational mass. This highlights the fact that we expect to be able to infer physical quantities from the observed neutrino signal. Finally, we  observed a characteristic correlation between the height of the luminosity peak during the plateau phase and its timing, a key prediction of modern CCSN theory. 

When examining the total energy radiated in neutrinos across all species, once again we found a monotonic relationship between progenitor compactness and the total energy radiated in neutrinos, with the energy ranging by a factor of four from the least-compact to the most-compact models. We also found that the relative species contribution of the total radiated energy is distinct for BH-forming models compared to the NS-formers. For the average neutrino energy as a function of time, while the $\nu_e$ energy fluctuates around a flat mean of 12.5 MeV for all of the models and the $\bar{\nu}_e$ and $\nu_{\mu}$ energies peak at 16-18 MeV and gradually decrease with time, the non-exploding, BH-forming models have a clear, distinguishable average energy evolution that rises faster and attain higher energies than even the most compact models. 

We then characterized the angular anisotropy of the neutrino emission. We found that the total energy radiated in neutrinos can vary significantly as a function of viewing angle, with the anisotropy of each model increasing with time, and more compact models having wider angular distributions. 

When estimating the detectability of various features in the luminosity signal, we found, as expected, that at a distance of 10 kiloparsecs, each of our models are expected to be detectable with event rates on the order of $10^3\ {\rm s}^{-1}$ to $10^6\ {\rm s}^{-1}$ across the JUNO, DUNE, SK, and IceCube detectors, enabling us to identify various aspects of the CCSN process, including the accretion-powered peak and the decay due to cooling. The compactness-dependent nesting is also maintained with each detector's event rate, with the rates varying by around a factor of four between the most- and least-compact progenitors. We also examined the effect of adiabatic MSW neutrino oscillation assuming a normal and inverted mass hierarchy, and found that while the oscillation models affected the overall event rates, the general patterns and features of the neutrino signal are maintained. 

Finally, we highlighted the correlation between the evolution of the radiated neutrino energy across all species and the GW f-mode. Since the evolution of the GW f-mode is closely tied to the equation-of-state governing the PNS \cite{Morozova2018}, due to this neutrino/GW correlation, the radiated neutrino energy may also help constrain the PNS nuclear equation-of-state.

Overall, there are many physical quantities and aspects of the CCSN process that can be inferred from the CCSN neutrino signal and which can serve as specific scientific targets for the neutrino detector community. From a CCSN neutrino signal, we have the potential to verify and/or glean information concerning:
\begin{enumerate}
    \item The main CCSN stages including the breakout burst, decay from the breakout burst, rise to peak, and Kelvin-Helmholtz cooling
    \item The progenitor core structure/compactness (from the luminosity peak height, total energy radiated, width of the breakout burst peak, and height of the early neutronization peak)
    \item The final NS gravitational mass (from correlations with the luminosity peak height and total radiated energy)
    \item Non-exploding BH formation (from the late-time luminosity signal, average energy evolution, and relative species composition of the total radiated energy)
    \item General BH formation (from the abrupt signal cutoff, relative neutrino species composition of the total radiated energy, and total radiated neutrino energy)
    \item The NS equation of state (from the total energy radiated in relation to the evolution of the GW f-mode)
\end{enumerate}

We leave an in-depth investigation into detector signals-to-noise and a discussion of the scientific advantages of combining detector signals to optimally extract physical information for future work. Moreover, there remains much work to be done concerning the effect of initial progenitor rotation, magnetic fields, and different nuclear equation-of-states on the neutrino signal. Nevertheless, this paper is meant in part to help the detector community move beyond simple demonstrations of detectability and a narrow focus on distinguishing the models of different groups to embracing the potential of supernova neutrino detection to usefully constrain the core astrophysics of supernova explosions.  

\section*{Data Availability}
The neutrino luminosity data for each model and species from this paper are available at \url{http://www.astro.princeton.edu/~burrows/nu-emissions.3d.update/} and \url{https://dvartany.github.io/data/} and the gravitational-wave data can be found at \url{http://www.astro.princeton.edu/~burrows/gw.3d.2024.update/}.

\bigskip     

\section*{Acknowledgments}
We thank Tianshu Wang for our long-term productive collaboration and his participation in the generation of many of the models used in this paper. AB acknowledges former support from the U.~S.\ Department of Energy Office of Science and the Office of Advanced Scientific Computing Research via the Scientific Discovery through Advanced Computing (SciDAC4) program and Grant DE-SC0018297 (subaward 00009650) and former support from the U.~S.\ National Science Foundation (NSF) under Grant AST-1714267. We are happy to acknowledge access to the Frontera cluster (under awards AST20020 and AST21003). This research is part of the Frontera computing project at the Texas Advanced Computing Center \citep{Stanzione2020}. Frontera is made possible by NSF award OAC-1818253. Additionally, a generous award of computer time was provided by the INCITE program, enabling this research to use resources of the Argonne Leadership Computing Facility, a DOE Office of Science User Facility supported under Contract DE-AC02-06CH11357. Finally, the authors acknowledge computational resources provided by the high-performance computer center at Princeton University, which is jointly supported by the Princeton Institute for Computational Science and Engineering (PICSciE) and the Princeton University Office of Information Technology, and our continuing allocation at the National Energy Research Scientific Computing Center (NERSC), which is supported by the Office of Science of the U.~S.\ Department of Energy under contract DE-AC03-76SF00098. 

\appendix
\section{SNEWPY Implementation}
While SNEWPY contains a large library of 1D, 2D, and 3D models developed by a variety of groups \cite{snewpy}, the models studied in this paper were developed recently and, as a result, were not included in this built-in library. Therefore, in order to plot the results of Figure \ref{detector_plots}, we modified a version of the SNEWPY code to be compatible with our new models, based on the existing SNEWPY infrastructure for previous F{\sc{ornax}} models. This involved writing a new class to the scripts \texttt{ccsn.py} and \texttt{ccsn\textunderscore loaders.py} (which we named ``Fornax\textunderscore2024'') that contained methods which pointed to our new models and reconstructed their luminosity profiles.

F{\sc{ornax}} uses 12 logarithmically-spaced energy bins, covering 1-300 MeV for the $\nu_e$ species and 1-100 MeV for the $\bar{\nu}_e$ and $\nu_{\mu}$ species \cite{Skinner2019}. This energy grid was then mapped to the SNOwGLoBES energy grid and configured into the form of a SNOwGLoBES input file using the SNEWPY module \texttt{generate\textunderscore fluence} in the \texttt{snewpy.snowglobes} script \cite{snewpy}. 

In order to model the event rate in a given detector, we used the module \texttt{simulate} in the  \texttt{snewpy.snowglobes} script for the specified detector. For example, for the SK detector, we employed the water Cherenkov configuration \texttt{wc100kt30prct} and scaled the final event rate by 0.325 to account for its 32.5 kilotonne volume.

\bibliography{thesis_refs}

\end{document}